\documentclass[leqno]{amsart}

\usepackage[foot]{amsaddr}

\usepackage{color}
\usepackage{geometry}
\usepackage{subfiles}
\usepackage{graphicx}
\usepackage[section]{placeins}
\usepackage{hyperref}
\usepackage{stmaryrd}
\usepackage{amsmath,amssymb,amsfonts,amsthm,textcomp}
\usepackage{mathtools}
\usepackage{tensor}
\usepackage{multicol}
\usepackage{subfig}
\usepackage{enumitem}
\usepackage{float}
\usepackage{booktabs}
\usepackage{mathbbol}
\usepackage{tikz}

\graphicspath{
              {./figures/}
             }

\newfont{\tenbfsl}{cmbxti9 scaled 1200}
\newfont{\tenbbb}{msbm10}
\newfont{\svnbbb}{msbm8}

\newcommand{\bs}[1]{\boldsymbol{#1}}
\newcommand{\br}[1]{\boldsymbol{\mathrm{{#1}}}}
\newcommand{\cl}[1]{\mathcal{#1}}

\newcommand{\jump}[1]{\llbracket {#1} \rrbracket}
\newcommand{\avg}[1]{\langle\mskip-4.5mu\langle {#1} \rangle\mskip-4.5mu\rangle}

\newcommand{\fr}[2]{\textstyle{\frac{{#1}}{{#2}}}}

\newcommand{\twovdots}{\mskip+2mu\colon\mskip-2mu}
\makeatletter
\def\threevdots{\mskip+4mu\vbox{\baselineskip2.25\p@ \lineskiplimit\z@
  \kern4.9\p@\hbox{.}\hbox{.}\hbox{.}}\mskip+3.8mu}
\makeatother

\newcommand{\dv}{\,\mathrm{d}v}
\newcommand{\da}{\,\mathrm{d}a}

\newcommand{\bdy}{\cl{R}}
\newcommand{\dbdy}{\partial\cl{R}}
\newcommand{\prt}{\cl{P}}
\newcommand{\dprt}{\partial\cl{P}}
\newcommand{\srf}{\cl{S}}

\newcommand{\intP}{\int\limits_{\prt}}
\newcommand{\intdP}{\int\limits_{\dprt}}

\newcommand{\trans}{\scriptscriptstyle\mskip-1mu\top\mskip-2mu}

\newcommand{\tr}{\mathrm{tr}\mskip2mu}

\newcommand{\Grad}{\mathrm{grad}\mskip2mu}
\newcommand{\Div}{\mathrm{div}\mskip2mu}

\newcommand{\Grads}{\Grad_{\mskip-2mu\scriptscriptstyle\cl{S}}}

\newcommand{\ju}{\bs{\jmath}_{\mathrm{u}}}
\newcommand{\je}{\bs{\jmath}_{\varepsilon}}
\newcommand{\jv}{\bs{\jmath}_{\mathrm{v}}}
\newcommand{\jl}{\bs{\jmath}_{\ell}}

\newcommand{\vv}{\bs{\upsilon}_{\mathrm{v}}}
\newcommand{\vl}{\bs{\upsilon}_{\mskip-1mu\ell}}
\newcommand{\vm}{\bs{\upsilon}_{\mathrm{m}}}
\newcommand{\vel}{\bs{\upsilon}}

\newcommand{\Dv}{\br{D}_{\mathrm{v}}}
\newcommand{\Dl}{\br{D}_{\mskip-1mu\ell}}

\newcommand{\Wl}{\br{W}_{\!\ell}}

\definecolor{newred}{RGB}{180,20,5}

\definecolor{newgreen}{RGB}{1,129,30}

\definecolor{mybluei}{RGB}{28,138,207}
\definecolor{myblueii}{RGB}{131,197,231}

\renewcommand{\jl}{\bs{u}}%{\vel_{\mathrm{rel}}}
\newcommand{\ddt}{\frac{\text{d}}{\text{d}t}}
\newcommand{\Dt}[1]{\frac{\text{D}#1}{\text{D}t}}

\newcommand{\mysetminusD}{\hbox{\tikz{\draw[line width=0.6pt,line cap=round] (3pt,0) -- (0,6pt);}}}
\newcommand{\mysetminusT}{\mysetminusD}
\newcommand{\mysetminusS}{\hbox{\tikz{\draw[line width=0.45pt,line cap=round] (2pt,0) -- (0,4pt);}}}
\newcommand{\mysetminusSS}{\hbox{\tikz{\draw[line width=0.4pt,line cap=round] (1.5pt,0) -- (0,3pt);}}}

\newcommand{\mysetminus}{\mathbin{\mathchoice{\mysetminusD}{\mysetminusT}{\mysetminusS}{\mysetminusSS}}}

\newcommand{\wall}{\mathcal{W}_t}

\begin{document}

\title[Compressible fluids with distinct mass and linear-momentum transport]{Compressible fluids with distinct mass and linear-momentum transport}

\author{Luis Espath$^\sharp$ \& Eliot Fried$^\flat$}

\address{$^\sharp$School of Mathematical Sciences\\
University of Nottingham\\
Nottingham, NG7 2RD, United Kingdom}
\email{luis.espath@nottingham.ac.uk}

\address{$^\flat$Mechanics and Materials Unit\\
Okinawa Institute of Science and Technology\\
Okinawa, Japan 904-0495}
\email{eliot.fried@oist.jp}

\date{\today}

\begin{abstract}
We formulate a thermodynamically consistent continuum theory for compressible, viscous, heat-conducting fluids in which the velocity entering the balance of mass is distinguished from the specific linear momentum entering the balances of linear momentum and energy. Starting from balances of mass, linear momentum, angular momentum, and internal energy, together with a power identity and the Clausius--Duhem inequality, we derive the mechanical and thermodynamic consequences of allowing these fields to differ. From local angular-momentum balance, we show that the Cauchy stress need not be symmetric and we determine its skew part. From the dissipation inequality, we obtain an admissible internal-energy flux and a closure in which the relative transport between mass and linear momentum is proportional to the pressure gradient rather than to the mass-density gradient. We also derive a free-enthalpy imbalance across shocks and a reduced wall dissipation inequality for rigid, impermeable walls undergoing prescribed rigid motion, together with simple admissible wall laws for temperature-controlled and heat-flow-controlled settings. For ideal gases, we write the governing equations in conservative dimensionless form, recover the classical compressible Navier--Stokes--Fourier theory when relative transport vanishes, and identify a distinguished low-Mach regime in which mass transport and linear-momentum transport remain distinct at leading order.
\end{abstract}

\maketitle


\vspace{0.5cm}

\textbf{AMS subject classifications:}
$\cdot$
76N99 % None of the above, but in this section - Compressible fluids and gas dynamics, general
$\cdot$
80A17 % Classical thermodynamics, heat transfer - Thermodynamics of continua
$\cdot$
82C26 % Statistical mechanics, structure of matter - Dynamic and nonequilibrium phase transitions (general)
$\cdot$
35L65 % Partial differential equations - Conservation laws
$\cdot$

\tableofcontents                        % Print table of contents

%-------------------------------------------------------------------------------%

\section{Introduction}

In the classical theory for the motion of a compressible, viscous, heat-conducting fluid---often specialized to the Navier--Stokes--Fourier equations---the balances of mass, linear momentum, and energy are formulated in terms of a single velocity field $\bs{\upsilon}$, together with the mass density $\varrho$ and the absolute temperature $\vartheta$. In that setting, the viscous contribution to the Cauchy stress depends on the symmetric part of $\Grad\bs{\upsilon}$, the heat flux is proportional to $\Grad\vartheta$, and closure is achieved by selecting independent constitutive variables together with a thermodynamic potential from which the remaining thermodynamic fields are obtained. Dissipative material coefficients, such as the shear and bulk viscosities and the thermal conductivity, are then specified in terms of the independent constitutive variables.

Brenner \cite{Bre05a,Bre05b} proposed a different kinematical setting involving two fields, a ``mass velocity'' $\vm$ and a ``volume velocity'' $\vv$. In this formulation, the balance of mass and the inertial terms are written with $\vm$, whereas the dissipative contribution to the Cauchy stress depends on the symmetric part of $\Grad\vv$. The energy flux combines the heat flux with a pressure-transport term involving $\vv-\vm$, and the classical Navier--Stokes--Fourier theory is recovered when $\vm\equiv\vv$. Because mass transport, momentum density, and dissipation are attached to different fields, however, the single velocity familiar from classical theory has no unique analogue in general.

In a later paper, Brenner \cite{Bre06} modified this framework by using the volume velocity $\vv$, rather than the mass velocity $\vm$, to form the linear-momentum and kinetic-energy densities, while leaving the constitutive dependence of the stress, the structure of the energy flux, and the wall data essentially unchanged. As shown in the Supplementary Material, that theory collapses to the Navier--Stokes--Fourier theory if either the balance of angular momentum or the second law of thermodynamics is imposed in addition to Brenner's \cite{Bre06} governing equations and constitutive prescriptions. 

Brenner \cite{Bre05a} also advances a conceptual rationale for distinguishing mass transport from volume transport: if volume is regarded as an extensive property of a fluid, then, like energy, entropy, or species mass, it should admit a balance law with advective, diffusive, and production terms. In a companion paper, Brenner \cite{Bre05b} supplements this viewpoint with empirical and kinetic-theory motivations, drawing attention to thermophoretic and related phoretic motions and arguing that Burnett and Maxwell thermal-stress effects may contribute at leading order when the asymptotics are organized by Mach number rather than Knudsen number. Interest in such ideas persists because they bear directly on compressible flows with strong density gradients, for which the classical Navier--Stokes--Fourier equations are known to be inadequate. Greenshields and Reese \cite{Gre07}, for example, used shock-wave structure as an independent test and reported improved predictions of shock thickness and density profiles, while Reddy et al.~\cite{Red19} and Reddy \& Dadzie \cite{Red20} reformulated the Navier--Stokes--Fourier equations in a volume-velocity setting to assess the role of mass-volume diffusion in strong-gradient flows, again with improved shock-structure predictions in the intermediate-Knudsen regime. These developments make it natural to ask whether a theory with distinct mass and linear-momentum transport can retain the intended advantages of such models while remaining compatible with the fundamental balance laws and the second law of thermodynamics.

Guided by that question, we formulate a continuum theory in which the velocity $\vel$ governing mass transport is distinguished from the specific linear momentum $\vl$ governing momentum transport. The balance of angular momentum plays a decisive role: once the momentum density is built from $\vl$ while mass is transported by $\vel$, symmetry of the Cauchy stress can no longer be assumed a priori. Instead, local moment balance determines its skew part through the mismatch between these transport mechanisms or, equivalently, determines the additional term that must accompany the linear-momentum balance to eliminate the spurious moment present in Brenner's \cite{Bre06} formulation. The second law imposes further constitutive restrictions through the Clausius--Duhem inequality. In particular, it identifies an admissible internal-energy flux, selects constitutive equations for the dissipative fields, and leads to a closure for the relative field $\jl=\vl-\vel$ driven by the pressure gradient rather than the mass-density gradient. The resulting framework yields shock and wall conditions compatible with both angular-momentum balance and nonnegative dissipation and recovers the classical Navier--Stokes--Fourier theory when $\vl\equiv\vel$.

The remainder of the paper is organized as follows. In Section~\S\ref{sec:basic.laws}, the balances of mass, linear momentum, and angular momentum are formulated, together with the associated power identity. In Section~\S\ref{sec:thermo}, the balance of internal energy and the Clausius--Duhem inequality are presented. In Section~\S\ref{sec:constitutive.restrictions}, constitutive restrictions implied by the second law are derived. In Section~\S\ref{sec:total.energy.balance}, the theory is recast in conservative form through a balance of total energy. In Section~\S\ref{sec:shock.dissipation}, the free-enthalpy imbalance at a shock is developed. In Section~\S\ref{sec:bc}, a rigid, impermeable solid wall undergoing prescribed rigid motion is considered; the corresponding wall kinematics are introduced; the reduced wall dissipation inequality is derived; and simple admissible wall laws are formulated. In Section~\S\ref{sec:asymptotic}, the theory is specialized to ideal gases, the governing equations are written in conservative dimensionless form, and the Navier--Stokes--Fourier limit along with a distinguished low-Mach regime is examined. In Section~\S\ref{sec:summary}, we summarize the main results and their implications. The transport and divergence identities used in the localizations are collected in the Appendix.

For convenience, we summarize the core notation used throughout this work.
\begin{align*}
\varrho &:\ \text{mass density}, &
\vartheta &:\ \text{absolute temperature}, \\[3pt]
\vel &:\ \text{velocity}, &
\vl &:\ \text{specific linear momentum}, \\[3pt]
\je &:\ \text{internal energy flux}, &
\jl = \vl - \vel &:\ \text{relative velocity}, \\[3pt]
\bs{q} &:\ \text{heat flux}, &
e &:\ \text{specific total energy}, \\[3pt]
\bs{b} &:\ \text{external body force density}, &
r &:\ \text{heat supply density}, \\[3pt]
\br{T} &:\ \text{Cauchy stress tensor}, &
\br{S} &:\ \text{symmetric (dissipative) stress}, \\[3pt]
\Dl &:\ \text{stretch tensor based on }\vl, &
\Dl^{0} &:\ \text{deviatoric part of }\Dl, \\[3pt]
p &:\ \text{thermodynamic pressure}, &
\kappa &:\ \text{thermal conductivity}, \\[3pt]
\mu, \zeta &:\ \text{shear viscosity, bulk viscosity}, &
\iota &:\ \text{mass-momentum modulus}, \\[3pt]
\bs{1} &:\ \text{second-order identity tensor}, &
\Grad, \Div &:\ \text{gradient and divergence operators}, \\[3pt]
\otimes &:\ \text{dyadic product}, &
\wedge &:\ \text{wedge product}.
\end{align*}

\section{Mechanical laws}
\label{sec:basic.laws}

Throughout this section, we work in a spatial (Eulerian) description. The fluid occupies a region $\bdy$ with boundary $\dbdy$, and the balance laws are formulated over an arbitrary fixed control volume $\prt$ with boundary $\dprt$ and outward unit normal $\bs{n}$. The region $\bdy$ may be divided into open complementary subregions $\bdy^-$ and $\bdy^+$ by a propagating shock surface $\srf_t\subset\bdy$, across which the fields entering the balance laws may suffer finite jumps. We denote by $\bs{m}$ the unit normal to $\srf_t$, directed from $\bdy^-$ into $\bdy^+$, and by $\xi$ the scalar normal velocity of $\srf_t$. The local consequences of the balance laws are obtained from their integral statements by use of the transport relation and divergence identities collected in Appendix~\ref{sec:identities}. For this purpose, we restrict attention to control volumes $\prt$ chosen so that $\prt\cap\srf_t\neq\varnothing$. Localization then yields field equations holding on $\bdy\mysetminus\srf_t$ together with the corresponding jump conditions on $\srf_t$.

\subsection{Balance of mass}

Let $\varrho$ be the mass density and let $\vel$ be the velocity. The balance of mass over a fixed control volume $\prt$ is expressed by
\begin{equation}\label{eq:partwise.bulk.mass}
\ddt\intP\varrho\dv
=-\intdP\varrho\vel\cdot\bs{n}\da.
\end{equation}
From \eqref{eq:partwise.bulk.mass}, together with the transport and divergence identities collected in Appendix~\ref{sec:identities}, we obtain the field equation
\begin{equation}\label{eq:mass.balance.gen}
\Dt{\varrho}+\varrho\mskip2mu\Div\vel=0
\qquad\text{in }\bdy\mysetminus\srf_t
\end{equation}
and the jump condition
\begin{equation}\label{eq:mass.balance.shock}
\jump{\varrho(\xi-\vel\cdot\bs{m})}=0
\qquad\text{on }\srf_t,
\end{equation}
where $\jump{f}=f^+-f^-$ denotes the jump of a field $f$ across $\srf_t$, with $f^\pm$ being the restriction of $f$ to $\dprt^\pm\cap\srf_t$. Defining the normal mass flux relative to $\srf_t$ by
\begin{equation}\label{eq:mass.flux.def}
\jmath
\coloneqq
\varrho^+(\xi-\vel^+\cdot\bs{m})
=\varrho^-(\xi-\vel^-\cdot\bs{m}),
\end{equation}
we infer from \eqref{eq:mass.balance.shock} that $\jmath$ is continuous across $\srf_t$, even though the mass density and velocity may suffer finite jumps there; in particular, no mass is created or destroyed as the shock propagates.

\subsection{Balance of linear momentum and balance of angular momentum}

We depart from the classical formulation by distinguishing the transport of mass from that of linear momentum. In place of the velocity $\vel$, we introduce a specific linear momentum $\vl$, taken to be independent of $\vel$, and identify $-\varrho\mathrm{D}\vl/\mathrm{D}t$ as the corresponding inertial body-force density.

Introducing the Cauchy stress tensor $\br{T}$ and the noninertial body-force density $\bs{b}$, we adopt the balance of linear momentum over a fixed control volume $\prt$ in the form
\begin{equation}\label{eq:partwise.linear.momentum}
\ddt\intP \varrho\vl\dv
=
-\intdP (\varrho\vl\otimes\vel)\bs{n}\da
+\intdP \br{T}\bs{n}\da
+\intP \bs{b}\dv,
\end{equation}
and the balance of angular momentum over $\prt$ in the form
\begin{equation}\label{eq:partwise.angular.momentum}
\ddt\intP \bs{r}\wedge\varrho\vl\dv
=
-\intdP \bs{r}\wedge(\varrho\vl\otimes\vel)\bs{n}\da
+\intdP \bs{r}\wedge\br{T}\bs{n}\da
+\intP \bs{r}\wedge\bs{b}\dv,
\end{equation}
where $\bs{r}(\bs{x})=\bs{x}-\bs{x}_0$ is the vector directed from a chosen point $\bs{x}_0$, about which moments are taken, to each point $\bs{x}$ in $\prt$, and where the wedge product of two vectors $\bs{p}$ and $\bs{q}$ is defined by
\begin{equation}\label{eq:wedge.product}
\bs{p}\wedge\bs{q}
\coloneqq
\bs{p}\otimes\bs{q}-\bs{q}\otimes\bs{p}.
\end{equation}

Using the transport relation \eqref{eq:reynolds.transport.shock} with $\varphi=\vl$ and the divergence identity \eqref{eq:ui2} with $\br{A}=\br{T}$, together with the jump condition \eqref{eq:mass.balance.shock} and the definition \eqref{eq:mass.flux.def} of the mass flux $\jmath$, we rewrite \eqref{eq:partwise.linear.momentum} in the form
\begin{equation}\label{eq:partwise.linear.momentum.plus}
\intP\Big(\varrho\Dt{\vl}-\Div\br{T}-\bs{b}\Big)\dv
-\int\limits_{\prt\cap\srf_t}(\jmath\jump{\vl}+\jump{\br{T}}\bs{m})\da_t
=\bf0.
\end{equation}
To determine the local content of \eqref{eq:partwise.linear.momentum.plus}, we choose $\prt$ to be compactly supported about a point away from $\srf_t$, so that $\prt\cap\srf_t=\varnothing$, and \eqref{eq:partwise.linear.momentum.plus} reduces to
\begin{equation}
\label{eq:pwlmbP}
\intP\Big(\varrho\Dt{\vl}-\Div\br{T}-\bs{b}\Big)\dv=\bf0.
\end{equation}
Since $\prt$ is otherwise arbitrary, the integrand on the left-hand side of \eqref{eq:pwlmbP} must vanish away from $\srf_t$, and we obtain the field equation
\begin{equation}\label{eq:pointwise.linear.momentum}
\varrho\Dt{\vl}
=
\Div\br{T}+\bs{b}
\qquad
\text{in }\bdy\mysetminus\srf_t.
\end{equation}
Substituting \eqref{eq:pointwise.linear.momentum} into \eqref{eq:partwise.linear.momentum.plus}, we are left with
\begin{equation}
\label{eq:pwebS}
\int\limits_{\prt\cap\srf_t}(\jmath\jump{\vl}+\jump{\br{T}}\bs{m})\da_t=\bf0.
\end{equation}
Next, we choose $\prt$ so that $\prt\cap\srf_t$ contains an interior point of $\srf_t$. Because such a patch may be made arbitrarily small while remaining otherwise arbitrary, the integrand on the left-hand side of \eqref{eq:pwebS} must vanish pointwise on $\srf_t$, and we obtain the jump condition
\begin{equation}\label{eq:linear.momentum.balance.shock}
\jmath\jump{\vl}+\jump{\br{T}}\bs{m}=\bf0
\qquad
\text{on }\srf_t.
\end{equation}

Using the transport relation \eqref{eq:reynolds.transport.shock} with $\varphi=\vl\wedge\vel$, the divergence identity \eqref{eq:ui7} with $\bs{g}=\bs{r}$ and $\br{A}=\br{T}$, and the identities
\begin{equation}
\Dt{\bs{r}}=\vel,
\qquad
\Grad\bs{r}=\br{1},
\end{equation}
the balance of angular momentum \eqref{eq:partwise.angular.momentum} may be written in the form
\begin{equation}\label{eq:partwise.angular.momentum.extended}
\intP\Big(
\bs{r}\wedge
\Big(\varrho\Dt{\vl}-\Div\br{T}-\bs{b}\Big)
+\br{T}-\br{T}^{\trans}-\varrho\vl\wedge\vel
\Big)\dv
-\int\limits_{\prt\cap\srf_t}
\bs{r}\wedge(\jmath\jump{\vl}+\jump{\br{T}}\bs{m})\da_t
=\bf0.
\end{equation}
Invoking the field equation \eqref{eq:pointwise.linear.momentum} and the jump condition \eqref{eq:linear.momentum.balance.shock} for the balance of linear momentum, we find that \eqref{eq:partwise.angular.momentum.extended} reduces to
\begin{equation}\label{eq:partwise.angular.momentum.reduced}
\intP(\br{T}-\br{T}^{\trans}-\varrho\vl\wedge\vel)\dv
=\bf0.
\end{equation}
Since $\prt$ is otherwise arbitrary, the integrand on the left-hand side of \eqref{eq:partwise.angular.momentum.reduced} must vanish away from $\srf_t$, and we obtain the field equation
\begin{equation}\label{eq:pointwise.angular.momentum}
\br{T}-\br{T}^{\trans}
=
\varrho\vl\wedge\vel
\qquad
\text{in }\bdy\mysetminus\srf_t,
\end{equation}
showing that the Cauchy stress tensor $\br{T}$ need not be symmetric if, as assumed here, the mechanisms governing the transport of mass and linear momentum differ. However, using the definition \eqref{eq:wedge.product} of the wedge product, we see that \eqref{eq:pointwise.angular.momentum} is equivalent to the requirement that the difference $\br{T}-\varrho\vl\otimes\vel$ be symmetric:
\begin{equation}\label{eq:angular.momentum.symmetry}
\br{T}-\varrho\vl\otimes\vel
=
(\br{T}-\varrho\vl\otimes\vel)^{\trans}
\qquad
\text{in }\bdy\mysetminus\srf_t.
\end{equation}

\subsection{Power identity}

Consistent with the postulated forms \eqref{eq:partwise.linear.momentum} and \eqref{eq:partwise.angular.momentum} of the balances of linear and angular momentum, we associate the specific linear momentum $\vl$ with the specific kinetic energy $\tfrac12|\vl|^2$. Taking the scalar product of the pointwise balance of linear momentum \eqref{eq:pointwise.linear.momentum} with $\vl$, we obtain the identity
\begin{equation}\label{eq:kinetic.energy.rate}
\varrho\Dt{(\frac{1}{2}|\vl|^2)}
=\Div(\br{T}^{\trans}\vl)-\br{T}\twovdots\Grad\vl
+\bs{b}\cdot\vl
\qquad
\text{in }\bdy\mysetminus\srf_t.
\end{equation}
Analogously, writing $\avg{f}=\tfrac12(f^+ + f^-)$ for the average of a field $f$ across $\srf_t$ and taking the scalar product of
\eqref{eq:linear.momentum.balance.shock} with $\avg{\vl}$, we obtain the identity
\begin{equation}\label{eq:kinetic.energy.jump}
\jmath\jump{\tfrac{1}{2}|\vl|^2}=\avg{\br{T}}\bs{m}\cdot\jump{\vl}-\jump{\br{T}^{\trans}\vl}\cdot\bs{m}\qquad
\text{on }\srf_t.
\end{equation}
Integrating \eqref{eq:kinetic.energy.rate} over $\prt$ and using the transport relation \eqref{eq:reynolds.transport.shock} with $\varphi=\tfrac{1}{2}|\vl|^2$, the divergence identity \eqref{eq:ui1}$_2$ with $\bs{g}=\br{T}^{\trans}\vl$, and \eqref{eq:kinetic.energy.jump}, we arrive at the identity
\begin{equation}\label{eq:power.identity}
\intdP\br{T}\bs{n}\cdot\vl\da+\intP\bs{b}\cdot\vl\dv
=\intP\br{T}\twovdots\Grad\vl\dv
+\int\limits_{\prt\cap\srf_t}\avg{\br{T}}\bs{m}\cdot\jump{\vl}\da_t
+\ddt\intP\tfrac12\varrho|\vl|^2\dv
+\intdP\tfrac12\varrho|\vl|^2\vel\cdot\bs{n}\da.
\end{equation}
Considering \eqref{eq:power.identity}, we define the power expenditures associated with external and internal actions by
\begin{equation}\label{eq:generic.external.power}
W^{\mathrm{e}}(\prt)\coloneqq\intdP\br{T}\bs{n}\cdot\vl\da+\intP\bs{b}\cdot\vl\dv
\end{equation}
and
\begin{equation}\label{eq:generic.internal.power}
W^{\mathrm{i}}(\prt)\coloneqq\intP\br{T}\twovdots\Grad\vl\dv
+\int\limits_{\prt\cap\srf_t}\avg{\br{T}}\bs{m}\cdot\jump{\vl}\da_t,
\end{equation}
respectively. With these definitions, we interpret \eqref{eq:power.identity} as a balance between the power expended on $\prt$ by external agencies and the sum of the power expended within $\prt$ and the total rate of change of kinetic energy in $\prt$. For $\vl\equiv\vel$, the external power \eqref{eq:generic.external.power}, internal power \eqref{eq:generic.internal.power}, and the associated balance \eqref{eq:power.identity} reduce to their classical counterparts.

\section{Thermodynamic laws}
\label{sec:thermo}

Throughout this section, we continue to work under the standing assumptions of Section~\ref{sec:basic.laws}. In particular, the thermodynamic laws are formulated over arbitrary fixed control volumes $\prt$ in regions where the
relevant fields are smooth, with the possible presence of a propagating shock surface $\srf_t$ across which finite jumps may occur. The local consequences of these laws follow from their integral statements by use of the transport relation and the divergence identities collected in Appendix~\ref{sec:identities}.

\subsection{Balance of energy}

Introducing the specific internal energy $\varepsilon$, the internal body-couple stress $\bs{\Lambda}=-\bs{\Lambda}^{\trans}$, the internal energy flux $\je$, and the heat supply density $r$, we adopt the balance of energy over a fixed control volume $\prt$ in the form
\begin{equation}
\label{eq:partwise.total.energy.balance}
\ddt\intP\varrho(\varepsilon+\tfrac12|\vl|^2)\dv
=
-\intdP\varrho(\varepsilon+\tfrac12|\vl|^2)\vel\cdot\bs{n}\da
+W^{\mathrm e}(\prt)
+\intP\bs{\Lambda}\twovdots\Wl\dv
-\intdP\je\cdot\bs{n}\da
+\intP r\dv,
\end{equation}
where
\begin{equation}
\label{eq:Wl.def}
\Wl\coloneqq\tfrac12(\Grad\vl-(\Grad\vl)^{\trans\mskip2mu})
\end{equation}
denotes the skew part of the gradient of the specific linear momentum. The internal body-couple stress $\bs{\Lambda}$ is associated with angular-momentum transfer arising from the distinction between the velocity $\vel$ and the specific linear momentum $\vl$. The internal energy flux $\je$ is written as $\je=\bs{q}+\bs{\jmath}$, where $\bs{q}$ denotes the heat flux and $\bs{\jmath}$ is an additional contribution associated with the mismatch
\begin{equation}
\label{eq:mismatch}
\jl\coloneqq\vl-\vel
\end{equation}
between the velocity and the specific linear momentum. The appearance of the body-couple power term and the decomposition of the internal energy flux are standard in continuum theories with multiple velocities and in far-from-equilibrium thermodynamics.

Using the power identity \eqref{eq:power.identity} to eliminate the external power $W^{\mathrm e}(\prt)$ from \eqref{eq:partwise.total.energy.balance}, and applying the transport relation \eqref{eq:reynolds.transport.shock} with $\varphi=\varepsilon$ together with the divergence identity \eqref{eq:ui1}$_2$ with $\bs{g}=\je$, we rewrite \eqref{eq:partwise.total.energy.balance} in the equivalent form
\begin{equation}
\label{eq:partwise.internal.energy.balance.rewritten}
\intP\Big(
\varrho\Dt{\varepsilon}
-\br{T}\twovdots\Grad\vl
-\bs{\Lambda}\twovdots\Wl
+\Div\je
-r
\Big)\dv
-\int\limits_{\prt\cap\srf_t}(\jmath\jump{\varepsilon}+\avg{\br{T}}\bs{m}\cdot\jump{\vl}
-\jump{\je}\cdot\bs{m}
)\da_t
=0.
\end{equation}
To determine the local content of \eqref{eq:partwise.internal.energy.balance.rewritten}, we choose $\prt$ to be compactly supported about a point away from $\srf_t$, so that $\prt\cap\srf_t=\varnothing$, and \eqref{eq:partwise.internal.energy.balance.rewritten} reduces to
\begin{equation}
\label{eq:pwebP}
\intP\Big(
\varrho\Dt{\varepsilon}
-\br{T}\twovdots\Grad\vl
-\bs{\Lambda}\twovdots\Wl
+\Div\je
-r
\Big)\dv=0.
\end{equation}
Since $\prt$ is otherwise arbitrary, the integrand on the left-hand side of \eqref{eq:pwebP} must vanish away from $\srf_t$, and we obtain the field equation
\begin{equation}
\label{eq:pointwise.energy.balance}
\varrho\Dt{\varepsilon}
=
\br{T}\twovdots\Grad\vl
+\bs{\Lambda}\twovdots\Wl
-\Div\je
+r
\qquad
\text{in }\bdy\mysetminus\srf_t.
\end{equation}
Substituting \eqref{eq:pointwise.energy.balance} into \eqref{eq:partwise.internal.energy.balance.rewritten}, we are left with
\begin{equation}
\label{eq:pwebS.final}
\int\limits_{\prt\cap\srf_t}
(\jmath\jump{\varepsilon}
+\avg{\br{T}}\bs{m}\cdot\jump{\vl}
-\jump{\je}\cdot\bs{m})
\da_t=0.
\end{equation}
Next, we choose $\prt$ so that $\prt\cap\srf_t$ contains an interior point of $\srf_t$. Because such a patch may be made arbitrarily small while remaining otherwise arbitrary, the integrand on the left-hand side of \eqref{eq:pwebS.final} must vanish pointwise on $\srf_t$, and we obtain the jump condition
\begin{equation}
\label{eq:energy.balance.shock}
\jmath\jump{\varepsilon}
+\avg{\br{T}}\bs{m}\cdot\jump{\vl}
-\jump{\je}\cdot\bs{m}
=0
\qquad
\text{on }\srf_t.
\end{equation}

\subsection{Imbalance of entropy}

Assuming that the absolute temperature $\vartheta$ is positive, we employ the conventional expressions $\bs{q}/\vartheta$ and $r/\vartheta$ for the entropy flux and entropy supply density, respectively, and adopt the Clausius--Duhem entropy imbalance over a control volume $\prt$ in the standard form
\begin{equation}\label{eq:entropy.imbalance}
\ddt\intP\varrho\eta\dv
\ge
-\intdP\varrho\eta\vel\cdot\bs{n}\da
-\intdP\frac{1}{\vartheta}\bs{q}\cdot\bs{n}\da
+\intP\frac{r}{\vartheta}\dv.
\end{equation}
Using the transport relation \eqref{eq:reynolds.transport.shock} with $\varphi=\eta$ and the divergence identity \eqref{eq:ui1}$_2$ with $\bs{g}=\bs{q}/\vartheta$, we rewrite \eqref{eq:entropy.imbalance} in the equivalent form
\begin{equation}\label{eq:entropy.imbalance.reworked}
\intP\frac{1}{\vartheta}\Big(\varrho\vartheta\Dt{\eta}+\Div\bs{q}-\bs{q}\cdot\Grad\ln\vartheta-r\Big)\dv
-\int\limits_{\prt\cap\srf_t}\Big(\jmath\jump{\eta}-\Big\llbracket\dfrac{\bs{q}}{\vartheta}\Big\rrbracket\cdot\bs{m}\Big)\da_t\ge 0.
\end{equation}
To determine the local content of \eqref{eq:entropy.imbalance.reworked}, we choose $\prt$ to be compactly supported about a point away from $\srf_t$, so that $\prt\cap\srf_t=\varnothing$, and obtain
\begin{equation}
\label{eq:pweimbP}
\intP\frac{1}{\vartheta}\Big(\varrho\vartheta\Dt{\eta}+\Div\bs{q}-\bs{q}\cdot\Grad\ln\vartheta-r\Big)\dv\ge 0.
\end{equation}
Since $\vartheta>0$ and $\prt$ is otherwise arbitrary, the integrand on the left-hand side of \eqref{eq:pweimbP} must be nonnegative away from $\srf_t$, and we obtain the field inequality
\begin{equation}\label{eq:entropy.imbalance.field.pre}
\varrho\vartheta\Dt{\eta}+\Div\bs{q}-\bs{q}\cdot\Grad\ln\vartheta-r\ge 0
\qquad
\text{in }\bdy\mysetminus\srf_t.
\end{equation}
Substituting \eqref{eq:entropy.imbalance.field.pre} into \eqref{eq:entropy.imbalance.reworked}, we are left with
\begin{equation}
\label{eq:pweimbS}
-\int\limits_{\prt\cap\srf_t}\Big(\jmath\jump{\eta}-\Big\llbracket\dfrac{\bs{q}}{\vartheta}\Big\rrbracket\cdot\bs{m}\Big)\da_t\ge 0.
\end{equation}
Next, we choose $\prt$ so that $\prt\cap\srf_t$ contains an interior point of $\srf_t$. Because such a patch may be taken arbitrarily small while remaining otherwise arbitrary, the integrand must satisfy
\begin{equation}\label{eq:entropy.imbalance.shock}
\jmath\jump{\eta}-\Big\llbracket\dfrac{\bs{q}}{\vartheta}\Big\rrbracket\cdot\bs{m}\le 0
\qquad
\text{on }\srf_t.
\end{equation}
Invoking the field equation \eqref{eq:pointwise.energy.balance} for the balance of internal energy to eliminate the heat supply density $r$ from \eqref{eq:entropy.imbalance.field.pre} and using the positivity of $\vartheta$, we obtain the local dissipation inequality
\begin{equation}\label{eq:pointwise.entropy.imbalance}
\varrho\Big(\Dt{\varepsilon}-\vartheta\Dt{\eta}\Big)-\br{T}\twovdots\Grad\vl+\bs{q}\cdot\Grad\ln\vartheta
-\bs{\Lambda}\twovdots\Wl
+\Div(\je-\bs{q})\le 0
\qquad
\text{in }\bdy\mysetminus\srf_t.
\end{equation}

To proceed, we introduce the specific free enthalpy
\begin{equation}\label{eq:free.enthalpy.def}
\chi\coloneqq\varepsilon-\vartheta\eta
+\frac{p}{\varrho},
\end{equation}
where $p$ denotes the thermodynamic pressure. In view of the local balance of angular momentum \eqref{eq:pointwise.angular.momentum}, the Cauchy stress tensor admits the representation
\begin{equation}\label{eq:stress.decomposition}
\br{T}=-p\bs{1}+\br{S}+\fr{1}{2}\varrho \mskip1mu\vl\wedge\vel,
\qquad
\br{S}=\br{S}^{\trans}.
\end{equation}
Since $p$ is the thermodynamic pressure, it is understood to be nondissipative. The spherical part of $\br{T}$ nevertheless contains, in general, the additional contribution $\tfrac13\tr\br{S}$, so that $\tfrac13\tr\br{T}=-p+\tfrac13\tr\br{S}$. Accordingly, the spherical contribution to $\br{T}$ may include a dissipative contribution through $\tr\br{S}$. Substituting \eqref{eq:free.enthalpy.def} and \eqref{eq:stress.decomposition} into \eqref{eq:pointwise.entropy.imbalance}, employing the identity $p\mskip2mu\Div\vl=p\mskip2mu\Div\vel
+\Div(p\jl)-\jl\cdot\Grad p$, and introducing the symmetric part,
\begin{equation}
\Dl\coloneqq\tfrac12(\Grad\vl+(\Grad\vl)^{\trans\mskip2mu}),
\end{equation}
of the gradient of the specific linear momentum $\vl$, we find that the dissipation inequality can be written in the form
\begin{equation}\label{eq:free.enthalpy.imbalance.pre}
\varrho\Big(\Dt{\chi}-\frac{1}{\varrho}\Dt{p}
+\eta\Dt{\vartheta}\Big)
-\br{S}\twovdots\Dl
+\bs{q}\cdot\Grad\ln\vartheta
-\jl\cdot\Grad p
-(\bs{\Lambda}+\fr{1}{2}\varrho \mskip1mu\vl\wedge\vel)\twovdots\Wl
+\Div(\je+p\jl-\bs{q})
\le 0.
\end{equation}

\section{Constitutive restrictions}
\label{sec:constitutive.restrictions}

The local balances \eqref{eq:mass.balance.gen}, \eqref{eq:pointwise.linear.momentum}, \eqref{eq:pointwise.angular.momentum}, and \eqref{eq:pointwise.energy.balance} for mass, linear momentum, angular momentum, and energy do not constitute a closed system of evolution equations: the number of fields entering these balances exceeds the number of governing equations. Completion of the description therefore necessitates the specification of additional constitutive relations. Apart from the first term on the left-hand side of \eqref{eq:free.enthalpy.imbalance.pre} and the divergence term of \eqref{eq:free.enthalpy.imbalance.pre}, each remaining contribution appears as a contraction between a constitutive quantity and an independently specifiable kinematic or thermodynamic process direction. Such contractions identify potential sources of dissipation and therefore constrain admissible constitutive dependence. In keeping with the constitutive structure of a simple fluid, we do not admit independent constitutive dependence on the skew part $\Wl$ of $\Grad\vl$. Accordingly, the constitutive mappings for $\chi$, $\eta$, $\varrho$, $\br{T}$, $\bs{\Lambda}$, $\bs q$, $\je$, and $\jl$ are taken to depend on $p$, $\vartheta$, $\Grad p$, $\Grad\vartheta$, and $\Dl$, but not on $\Wl$.

Recognizing that the dissipation inequality must hold for all admissible local processes consistent with the balance laws, we consider processes for which $\Dl$, $\Wl$, $\Grad p$, $\Grad\vartheta$, and $\jl$ vanish at a given point and time, while $\je+p\jl-\bs q$ is not spatially uniform. For such processes, all remaining contributions to \eqref{eq:free.enthalpy.imbalance.pre} vanish identically, and the inequality reduces locally to $\Div(\je+p\jl-\bs q)\le0$. Since $\je+p\jl-\bs q$ is otherwise unrestricted, its divergence may assume either sign, and \eqref{eq:free.enthalpy.imbalance.pre} can therefore be violated unless this divergence vanishes identically. We therefore conclude that admissibility of the dissipation inequality for all processes requires $\je+p\jl-\bs q$ to be divergence-free. Because the addition of a divergence-free contribution to the internal energy flux does not affect the transfer of energy, this requirement is satisfied by imposing the constitutive restriction
\begin{equation}\label{eq:CN.energy.flux.restriction}
\je\coloneqq \bs q-p\jl,
\end{equation}
up to the addition of the curl of a vector field. Any such contribution with identically vanishing divergence would leave the local bulk dissipation inequality unchanged, but would require separate treatment at shocks and boundaries.

To treat the term involving $\Wl$ in \eqref{eq:free.enthalpy.imbalance.pre}, we next consider admissible processes for which $\Dl$, $\Grad p$, $\Grad\vartheta$, and the material derivatives of $p$ and $\vartheta$ vanish at a given point and time, while $\Wl$ is prescribed arbitrarily. For such processes, all remaining contributions to \eqref{eq:free.enthalpy.imbalance.pre} vanish at that point. Since the inequality must hold for both $\Wl$ and $-\Wl$, it follows that
$\bs{\Lambda}+\tfrac12\varrho\vl\wedge\vel=\bs0$,
and, thus, that
\begin{equation}\label{eq:CN.lambda.restriction}
\bs{\Lambda}=\tfrac12\vel\wedge\varrho\vl.
\end{equation}

With the restrictions \eqref{eq:CN.energy.flux.restriction} and \eqref{eq:CN.lambda.restriction} in place, the dissipation inequality \eqref{eq:free.enthalpy.imbalance.pre} reduces to
\begin{equation}\label{eq:reduced.entropy.imbalance}
\varrho\Big(\Dt{\chi}-\frac{1}{\varrho}\Dt{p}
+\eta\Dt{\vartheta}\Big)
-\br S\twovdots\Dl
+\bs q\cdot\Grad\ln\vartheta
-\jl\cdot\Grad p
\le 0.
\end{equation}
Recalling the decision to treat the specific free enthalpy $\chi$, the mass density $\varrho$, the entropy $\eta$, the symmetric extra stress $\br S$, the heat flux $\bs q$, and the relative velocity $\jl$ as constitutive response functions of the fields $(p,\vartheta,\Grad p,\Grad\vartheta,\Dl)$, we apply the conventional Coleman--Noll argument. Since the material derivatives of $p$ and $\vartheta$ may be prescribed independently at a point, \eqref{eq:reduced.entropy.imbalance} can be violated unless the response functions $\hat\chi$, $\hat\varrho$, and $\hat\eta$ for $\chi$, $\varrho$, and $\eta$ are independent of $(\Grad p,\Grad\vartheta,\Dl)$ and satisfy the classical thermodynamic relations
\begin{equation}\label{eq:CN.thermodynamic.relations}
\hat\varrho(p,\vartheta)=\frac{1}{\partial_p\hat\chi(p,\vartheta)}
\qquad\text{and}\qquad
\hat\eta(p,\vartheta)=-\partial_\vartheta\hat\chi(p,\vartheta),
\end{equation}
where subscripts denote partial differentiation with respect to $p$ and $\vartheta$. Consequently, \eqref{eq:free.enthalpy.imbalance.pre} reduces further to the residual dissipation inequality
\begin{equation}\label{eq:residual.dissipation.inequality}
-\br{S}\twovdots\Dl
+\bs{q}\cdot\Grad\ln\vartheta
-\jl\cdot\Grad p
\le 0.
\end{equation}
To ensure satisfaction of \eqref{eq:residual.dissipation.inequality}, we restrict attention to constitutive prescriptions of the form
\begin{equation}\label{eq:constitutive.prescriptions.S.pi.q}
\br{S} = 2\mu(p,\vartheta)\Dl^{0}
+\zeta(p,\vartheta)\tr(\Dl)\bs{1},
\qquad
\bs{q}=-\kappa(p,\vartheta) \mskip2mu \Grad\vartheta,
\qquad
\jl=\iota(p,\vartheta) \mskip2mu \Grad p,
\end{equation}
where the shear viscosity $\mu$, bulk viscosity $\zeta$, thermal conductivity $\kappa$, and mass--momentum modulus $\iota$ satisfy
\begin{equation}
\mu(p,\vartheta)\ge0,
\qquad
\zeta(p,\vartheta)\ge0,
\qquad
\kappa(p,\vartheta)\ge0,
\qquad\text{and}\qquad
\iota(p,\vartheta)\ge0,
\end{equation}
and $\Dl^{0}\coloneqq\Dl-\tfrac{1}{3}(\tr\Dl)\bs{1}$ denotes the deviatoric part of $\Dl$.

\section{Balance of total energy}
\label{sec:total.energy.balance}

Defining the specific total energy $e$ by
\begin{equation}
\label{eq:total.specific.energy}
e \coloneqq \varepsilon + \tfrac12 |\vl|^2
\end{equation}
and invoking the final expression
\eqref{eq:generic.external.power} for the external power $W^{\mathrm e}(\prt)$ together with the
constitutive restriction \eqref{eq:CN.lambda.restriction} on $\bs{\Lambda}$, we find that the original version \eqref{eq:partwise.total.energy.balance} of the balance of energy for $\prt$ can be expressed as
\begin{equation}\label{eq:partwise.total.energy.balance.final}
\ddt\intP \varrho e \dv
=
-\intdP \varrho e\vel\cdot\bs{n}\da
+\intdP(\br{T}^{\trans}\vl-\je)\cdot\bs{n}\da
+\intP(
r
+ \bs{b}\cdot\vl
+ \tfrac12(\varrho \mskip1mu \vel\wedge\vl)\twovdots\Wl
)\dv.
\end{equation}
Localizing \eqref{eq:partwise.total.energy.balance.final}, we obtain a corresponding field equation 
\begin{equation}\label{eq:pointwise.total.energy.balance}
\varrho \Dt{e}
-\Div(\br{T}^{\trans}\vl-\je)
+\tfrac12(\varrho \mskip1mu \vl\wedge\vel)\twovdots\Wl
-r
-\bs{b}\cdot\vl
=0.
\end{equation}

\section{Dissipation inequality at a shock}
\label{sec:shock.dissipation}

The jump conditions associated with the balances of mass, the balance of linear momentum, the balance of energy, and the imbalance of entropy are given by \eqref{eq:mass.balance.shock}, \eqref{eq:linear.momentum.balance.shock}, \eqref{eq:energy.balance.shock}, and \eqref{eq:entropy.imbalance.shock}, respectively. For convenience, we collect them here:
\begin{equation}\label{eq:shock.jump.system}
\left\{
\begin{aligned}
&\jump{\jmath} = 0, \\[4pt]
&\jmath \jump{\vl} + \jump{\br{T}} \bs{m}=\bf0, \\[4pt]
&\jmath \jump{\varepsilon}
+\jump{\vl} \cdot \avg{\br{T}} \bs{m}
-\jump{\je} \cdot \bs{m}=0, \\[4pt]
&\jmath \jump{\eta}
-\Big\llbracket\dfrac{\bs{q}}{\vartheta}\Big\rrbracket \cdot \bs{m}\le0.
\end{aligned}
\right.
\end{equation}
To isolate the mechanical and thermal contributions to dissipation at $\srf_t$, we multiply \eqref{eq:shock.jump.system}$_4$ by the average $\avg{\vartheta}$ and subtract the resulting inequality from \eqref{eq:shock.jump.system}$_3$. This gives:
\begin{equation}\label{eq:shock.dissipation.inequality}
\jmath (\jump{\varepsilon}-\avg{\vartheta}\jump{\eta})
\ge
- \jump{\vl}\cdot\avg{\br{T}}\bs{m}
+\Big(\jump{\je}-\avg{\vartheta}\Big\llbracket\dfrac{\bs{q}}{\vartheta}\Big\rrbracket \Big)\cdot\bs{m}.
\end{equation}
Using the decomposition \eqref{eq:stress.decomposition} of $\br{T}$ and the stipulated restriction \eqref{eq:CN.energy.flux.restriction} on $\je$, and invoking the identity
\begin{equation}
\avg{p}\jump{\vl}\cdot\bs{m}
-\jump{p\jl}\cdot\bs{m}
=
-\jump{p}\avg{\jl}\cdot\bs{m}
+\avg{p}\jump{\vel}\cdot\bs{m},
\end{equation}
we rewrite \eqref{eq:shock.dissipation.inequality} in the form
\begin{equation}\label{eq:shock.dissipation.expanded}
\jmath (\jump{\varepsilon}-\avg{\vartheta}\jump{\eta})
\ge
-\jump{\vl}\cdot\avg{\br{S}+\tfrac{1}{2}\varrho \vl\wedge\vel}\bs{m}
-\jump{p}\avg{\jl}\cdot\bs{m}
+\Big(\jump{\je}-\avg{\vartheta}\Big\llbracket\dfrac{\bs{q}}{\vartheta}\Big\rrbracket \Big)\cdot\bs{m}
+\avg{p}\jump{\vel}\cdot\bs{m}.
\end{equation}
Introducing the perpendicular projector
\begin{equation}
\br{P}\coloneqq\bs{1}-\bs{m}\otimes\bs{m}
\end{equation}
onto $\srf_t$, and decomposing the traction into its normal and tangential components
\begin{equation}\label{eq:tm.tau}
t_m\coloneqq
\bs{m}\cdot\avg{\br{S}}\bs{m}
\qquad\text{and}\qquad
\bs{\tau}\coloneqq
\br{P}\avg{\br{S}+\tfrac{1}{2}\varrho \vl\wedge\vel}\bs{m},
\end{equation}
we find that \eqref{eq:shock.dissipation.expanded} can be written as
\begin{equation}\label{eq:shock.dissipation.final}
\jmath (\jump{\varepsilon}-\avg{\vartheta}\jump{\eta})
\ge
- t_m\,\jump{\vl}\cdot\bs{m}
- \bs{\tau}\cdot\jump{\vl}
-\jump{p}\avg{\jl}\cdot\bs{m}
+\Big(\jump{\je}-\avg{\vartheta}\Big\llbracket\dfrac{\bs{q}}{\vartheta}\Big\rrbracket \Big)\cdot\bs{m}
+\avg{p}\jump{\vel}\cdot\bs{m}.
\end{equation}
Furthermore, substituting the consequence
\begin{equation}
\jump{\vel\cdot\bs{m}}
=-\jump{\xi-\vel\cdot\bs{m}}
=-\Big\llbracket\dfrac{\varrho(\xi-\vel\cdot\bs{m})}{\varrho}\Big\rrbracket
=-\avg{\varrho(\xi-\vel\cdot\bs{m})}\Big\llbracket\dfrac{1}{\varrho}\Big\rrbracket
-\jump{\varrho(\xi-\vel\cdot\bs{m})}
\Big\langle\mskip-8mu\Big\langle\dfrac{1}{\varrho}\Big\rangle\mskip-8mu\Big\rangle
=-\jmath\Big\llbracket\dfrac{1}{\varrho}\Big\rrbracket,
\end{equation}
of the definition \eqref{eq:mass.flux.def} of $\jmath$ and \eqref{eq:shock.jump.system}$_1$ into \eqref{eq:shock.dissipation.final}, we find that
\begin{equation}\label{eq:shock.dissipation.normal.tangential.final}
\jmath\Big(\jump{\varepsilon}-\avg{\vartheta}\jump{\eta}
+\avg{p}\Big\llbracket\dfrac{1}{\varrho}\Big\rrbracket\Big)
\ge
- t_m\jump{\vl}\cdot\bs{m}
- \bs{\tau}\cdot\jump{\vl}
-\jump{p}\avg{\jl}\cdot\bs{m}
+\Big(\jump{\je}-\avg{\vartheta}\Big\llbracket\dfrac{\bs{q}}{\vartheta}\Big\rrbracket \Big)\cdot\bs{m}.
\end{equation}
Finally, noting that
\begin{equation}
\jump{\varepsilon}-\avg{\vartheta}\jump{\eta}
+\avg{p}\Big\llbracket\dfrac{1}{\varrho}\Big\rrbracket
=\Big\llbracket
\varepsilon-\vartheta\eta+\frac{p}{\varrho}\mskip2mu
\Big\rrbracket
+\avg{\eta}\jump{\vartheta}
-\Big\langle\mskip-8mu\Big\langle\dfrac{1}{\varrho}\Big\rangle\mskip-8mu\Big\rangle\jump{p} 
=\jump{\chi}
+\avg{\eta}\jump{\vartheta}
-\Big\langle\mskip-8mu\Big\langle\dfrac{1}{\varrho}\Big\rangle\mskip-8mu\Big\rangle\jump{p},
\end{equation}
we arrive at the free-enthalpy imbalance across $\srf_t$:
\begin{equation}\label{eq:shock.free.enthalpy.imbalance}
\jmath\Big(\jump{\chi}
-\jump{p}\Big\langle\mskip-8mu\Big\langle\dfrac{1}{\varrho}\Big\rangle\mskip-8mu\Big\rangle
+\jump{\vartheta}\avg{\eta}\Big)
\ge
- t_m\jump{\vl}\cdot\bs{m}
- \bs{\tau}\cdot\jump{\vl}
-\jump{p}\avg{\jl}\cdot\bs{m}
+\Big(\jump{\je}-\avg{\vartheta}\Big\llbracket\dfrac{\bs{q}}{\vartheta}\Big\rrbracket \Big)\cdot\bs{m}.
\end{equation}

In particular, if the absolute temperature $\vartheta$ is continuous across $\srf_t$, then \eqref{eq:shock.free.enthalpy.imbalance} specializes to
\begin{equation}\label{eq:shock.free.enthalpy.imbalance.2}
\jmath \Big(\jump{\chi}
-\jump{p}\Big\langle\mskip-8mu\Big\langle\dfrac{1}{\varrho}\Big\rangle\mskip-8mu\Big\rangle\Big)
\ge
- t_m\,\jump{\vl}\cdot\bs{m}
- \bs{\tau}\cdot\jump{\vl}
-\jump{p}\avg{\jl}\cdot\bs{m}.
\end{equation}

\section{Boundary conditions at a solid wall}
\label{sec:bc}

We consider a rigid, impermeable solid wall with boundary surface $\wall=\partial\bdy^{\mathrm{wall}}$. The unit normal to $\wall$ is denoted by $\bs{\nu}$ and is directed into the region occupied by the fluid. The wall is assumed to be free of superficial structure: there is no surface excess mass, energy, or entropy, no surface stress, and no surface heat flux. Accordingly, the only mechanisms by which the wall interacts with the fluid are through the classical traction and energy flux transmitted across $\wall$.

\subsection{Kinematics and essential wall data}

Letting $\bs{w}$ denote the prescribed wall velocity on $\wall$, we decompose it into normal and tangential parts according to
\begin{equation}
\label{eq:wall.velocity}
\bs{w}=w\bs{\nu}+\bs{w}_{\mathrm{tan}},
\qquad
\bs{w}_{\mathrm{tan}}\cdot\bs{\nu}=0,
\qquad
\text{on }\wall.
\end{equation}
For a rigid wall, $\bs{w}$ is taken to be the restriction to $\wall$ of a rigid-body velocity field.

Consistent with impermeability, we stipulate that the velocity $\vel$ satisfy
\begin{equation}
\label{eq:impermeability}
\vel\cdot\bs{\nu}=w
\qquad
\text{on }\wall.
\end{equation}
As an additional essential boundary prescription, we stipulate that the tangential component of the specific linear momentum $\vl$ satisfy
\begin{equation}
\label{eq:slip}
\br{P}\vl=\bs{w}_{\mathrm{tan}}
\qquad
\text{on }\wall,
\end{equation}
where
\begin{equation}
\br{P}\coloneqq\br{1}-\bs{\nu}\otimes\bs{\nu}
\end{equation}
is now, and hereafter, the orthogonal projector onto the tangent plane of $\wall$.

At this stage, no restriction is imposed on the tangential component
\begin{equation}
\vel_{\mathrm{tan}}\coloneqq\br{P}\vel
\end{equation}
of $\vel$ or on the normal component
\begin{equation}
V_\ell\coloneqq\vl\cdot\bs{\nu}
\end{equation}
of $\vl$ on $\wall$. Thus, the essential wall data fix the normal component of $\vel$ and the tangential component of $\vl$, while the complementary quantities $\vel_{\mathrm{tan}}$ and $V_\ell$ remain available for subsequent boundary prescriptions.

Using \eqref{eq:mismatch}, together with \eqref{eq:impermeability} and \eqref{eq:slip}, we find that the normal and tangential components of the mismatch $\jl$ are given by
\begin{equation}
\label{eq:u.wall.decomp}
\jl\cdot\bs{\nu}=V_\ell-w
\qquad
\text{and}
\qquad
\br{P}\jl=\bs{w}_{\mathrm{tan}}-\vel_{\mathrm{tan}}
\qquad
\text{on }\wall.
\end{equation}

\subsection{Boundary pillbox localization and reduced wall dissipation}
\label{sec:7.2}

We now adapt the boundary-pillbox argument to the present setting of a rigid, impermeable wall. Let $A\subset\wall$ be arbitrary, and let $P_h$ denote a thin pillbox obtained by offsetting $A$ a distance $h$ into the fluid and closing it by a lateral surface normal to the wall. We regard $P_h$ as attached to the wall and hence convecting with the prescribed wall velocity $\bs{w}$. Because the wall is free of superficial structure, there are no surface excess quantities, no surface stress, and no surface heat flux. Consequently, as $h\to0$, the only wall-side interaction terms that survive are the traction exerted by the wall on the fluid and the heat transmitted from the wall to the fluid.

Let $\bs{t}^{\mathrm{wall}}$ denote the traction exerted by the wall on the fluid, let $q^{\mathrm{wall}}$ denote the scalar heat flow from the wall into the fluid per unit area, and let $\vartheta^{\mathrm{wall}}$ denote the wall temperature. From the balances of linear momentum and energy, we obtain the conditions

\begin{equation}
\label{eq:wall.force.balance}
\bs{t}^{\mathrm{wall}}+\br{T}\bs{\nu}=\bs{0}
\qquad\text{and}\qquad
\bs{t}^{\mathrm{wall}}\cdot\bs{w}+\br{T}\bs{\nu}\cdot\vl+q^{\mathrm{wall}}-\je\cdot\bs{\nu}
=0.
\end{equation}
Similarly, from the imbalance of entropy, we obtain the condition
\begin{equation}
\label{eq:wall.entropy.imbalance.raw}
\frac{q^{\mathrm{wall}}}{\vartheta^{\mathrm{wall}}}-\frac{\bs{q}\cdot\bs{\nu}}{\vartheta}\le 0.
\end{equation}

Using \eqref{eq:stress.decomposition}$_1$, \eqref{eq:CN.energy.flux.restriction}, \eqref{eq:slip}, \eqref{eq:u.wall.decomp}$_1$, and \eqref{eq:wall.force.balance}$_1$, we find that \eqref{eq:wall.force.balance}$_2$ simplifies to
\begin{equation}
\label{eq:wall.energy.balance.reduced}
\bs{\nu}\cdot\br{S}\bs{\nu}(V_\ell-w)
+q^{\mathrm{wall}}-\bs{q}\cdot\bs{\nu}=0.
\end{equation}
Eliminating $q^{\mathrm{wall}}$ between \eqref{eq:wall.entropy.imbalance.raw} and \eqref{eq:wall.energy.balance.reduced}, we arrive at the reduced wall dissipation inequality
\begin{equation}
\label{eq:wall.dissipation}
\bs{\nu}\cdot\br{S}\bs{\nu}(V_\ell-w)+\frac{\vartheta^{\mathrm{wall}}-\vartheta}{\vartheta}\bs{q}\cdot\bs{\nu}\ge 0.
\end{equation}
Thus, once the essential wall data from Subsection~7.1 have been prescribed, the only mechanical contribution to entropy production at the wall is the scalar pairing between the normal component $V_\ell-w$ of the mismatch $\jl=\vl-\vel$ and the normal component of the dissipative traction $\bs{\nu}\cdot\br{S}\bs{\nu}$.

For later use, it is convenient to record the tangential traction
\begin{equation}
\label{eq:wall.tangential.traction}
\bs{\tau}\coloneqq\br{P}\br{T}\bs{\nu}=\br{P}\br{S}\bs{\nu}+\tfrac12\varrho\br{P}(\vl\wedge\vel)\bs{\nu}.
\end{equation}
Using \eqref{eq:wall.velocity}, \eqref{eq:slip}, and \eqref{eq:u.wall.decomp}, we find that
\begin{align}
\br{P}(\vl\wedge\vel)\bs{\nu}
=
\br{P}(\vl(\vel\cdot\bs{\nu})-\vel(\vl\cdot\bs{\nu}))
&=
w\bs{w}_{\mathrm{tan}}-V_\ell\vel_{\mathrm{tan}}
\notag\\[4pt]
&=
w(\bs{w}_{\mathrm{tan}}-\vel_{\mathrm{tan}})+(w-V_\ell)\vel_{\mathrm{tan}}
=
w\br{P}\jl-(\jl\cdot\bs{\nu})\vel_{\mathrm{tan}},
\label{eq:wall.wedge.identity}
\end{align}
and, accordingly, that
\begin{equation}
\bs{\tau}=\br{P}\br{S}\bs{\nu}+\tfrac12\varrho(w\br{P}\jl-(\jl\cdot\bs{\nu})\vel_{\mathrm{tan}}).
\end{equation}
Thus, although $\bs{\tau}$ does not enter the reduced dissipation inequality \eqref{eq:wall.dissipation}, it still depends on the quantities $\vel_{\mathrm{tan}}$ and $V_\ell$ that are available for prescription. In particular, \eqref{eq:slip} eliminates the tangential mismatch that would appear in the general wall dissipation inequality; accordingly, $\bs{\tau}$ enters our formulation as a reaction associated with the essential tangential prescription, rather than as the subject of an additional wall constitutive law.

\subsection{Simple wall laws}

Apart from the essential boundary conditions \eqref{eq:impermeability} and \eqref{eq:slip}, it remains to supplement the wall model by one scalar mechanical relation involving $V_\ell-w$ and $\bs{\nu}\cdot\br{S}\bs{\nu}$, together with one thermal relation. This viewpoint is consistent with the simple boundary-condition framework of Fried and Gurtin~\cite{Fri07}, according to which the relevant mechanical and thermal data are prescribed and the remaining wall fields are determined reactively from the balances.

For the mechanical part of \eqref{eq:wall.dissipation}, nondissipative wall behavior is obtained by enforcing equality. This may be achieved by imposing either
\begin{equation}
\label{eq:wall.mechanical.nondissipative}
V_\ell=w
\qquad
\text{or}
\qquad
\bs{\nu}\cdot\br{S}\bs{\nu}=0
\qquad
\text{on }\wall.
\end{equation}
A simple dissipative normal wall law is obtained by prescribing
\begin{equation}
\label{eq:wall.mechanical.dissipative}
\bs{\nu}\cdot\br{S}\bs{\nu}=\beta(V_\ell-w)
\qquad
\text{on }\wall,
\end{equation}
where $\beta$ may depend on $(p,\vartheta)$ and satisfies $\beta\ge0$. Thus,
\begin{equation}
\bs{\nu}\cdot\br{S}\bs{\nu}(V_\ell-w)=\beta(V_\ell-w)^2\ge0.
\end{equation}
The tangential traction $\bs{\tau}$ is treated here as a reaction associated with the essential tangential condition \eqref{eq:slip}. This is consistent with the power structure of the theory, for which the traction $\br{T}\bs{\nu}$ is power-conjugate to the trace of $\vl$, not to the trace of $\vel$. Consequently, the tangential mechanical term in the general wall dissipation involves $\br{P}\vl-\bs{w}_{\mathrm{tan}}$, not $\vel_{\mathrm{tan}}$. Once \eqref{eq:slip} has been imposed, that tangential mismatch vanishes identically, and no additional constitutive wall law for $\bs{\tau}$ is introduced. Instead, $\bs{\tau}$ is recovered, when needed, from the bulk solution and the prescribed wall data through \eqref{eq:wall.tangential.traction}. Correspondingly, $\vel_{\mathrm{tan}}$ is not an independently constitutively paired wall variable. By \eqref{eq:u.wall.decomp}$_2$, $\vel_{\mathrm{tan}}=\bs{w}_{\mathrm{tan}}-\br{P}\jl$ on $\wall$, and, using the constitutive relation for $\jl$, this quantity is determined through the bulk fields once the essential wall data have been prescribed.

If the wall temperature $\vartheta^{\mathrm{wall}}$ is prescribed, then \eqref{eq:wall.dissipation} is the relevant thermal restriction. Nondissipative thermal behavior is obtained by imposing
\begin{equation}
\vartheta=\vartheta^{\mathrm{wall}}
\qquad
\text{on }\wall.
\end{equation}
A simple dissipative choice is
\begin{equation}
\bs{q}\cdot\bs{\nu}=\gamma
(\vartheta^{\mathrm{wall}}-\vartheta)
\qquad
\text{on }\wall,
\end{equation}
where $\gamma$ may depend on $(p,\vartheta)$ and satisfies $\gamma\ge0$. More generally, any constitutive prescription for $\bs{q}\cdot\bs{\nu}$ satisfying
\begin{equation}
(\vartheta^{\mathrm{wall}}-\vartheta)\bs{q}\cdot\bs{\nu}\ge0
\qquad
\text{on }\wall
\end{equation}
is admissible. If desired, a quartic alternative may be used in place of the linear law.

If instead the wall heat flow $q^{\mathrm{wall}}$ is prescribed, then it is more natural to work with \eqref{eq:wall.entropy.imbalance.raw} and \eqref{eq:wall.energy.balance.reduced} rather than with \eqref{eq:wall.dissipation}. In that case, the normal component of the heat flux is determined by
\begin{equation}
\label{eq:wall.heatflow.balance}
\bs{q}\cdot\bs{\nu}=q^{\mathrm{wall}}+\bs{\nu}\cdot\br{S}\bs{\nu}(V_\ell-w),
\end{equation}
and we may eliminate $\bs{q}\cdot\bs{\nu}$ to arrive at the alternative reduced dissipation inequality
\begin{equation}
\label{eq:wall.dissipation.heatflow}
\bs{\nu}\cdot\br{S}\bs{\nu}(V_\ell-w)+\frac{\vartheta^{\mathrm{wall}}-\vartheta}{\vartheta^{\mathrm{wall}}}q^{\mathrm{wall}}\ge0.
\end{equation}
Here $\vartheta^{\mathrm{wall}}$ is not prescribed a priori but instead serves as a reactive wall temperature compatible with the imposed heat flow. The special case
\begin{equation}
\label{eq:wall.zero.heatflow}
q^{\mathrm{wall}}=0
\qquad
\text{on }\wall
\end{equation}
is a model for a thermally insulated wall. If, in addition, the mechanical contribution vanishes, then \eqref{eq:wall.heatflow.balance} reduces to $\bs{q}\cdot\bs{\nu}=0$. More generally, when the mechanical contribution vanishes, a positive prescribed wall heat flow is accompanied by the requirement that $\vartheta^{\mathrm{wall}}\ge\vartheta$, whereas a negative prescribed wall heat flow is accompanied by the requirement that $\vartheta^{\mathrm{wall}}\le\vartheta$.

\section{Ideal-gas specialization, dimensionless form, and asymptotic limits}
\label{sec:asymptotic}

We now specialize the bulk equations to ideal gases, write them in conservative dimensionless form, and identify two limiting regimes of particular interest. The first is the Navier--Stokes--Fourier limit, obtained when the relative transport of mass and linear momentum disappears. The second is a distinguished low-Mach regime in which that relative transport survives at leading order. The associated shock and wall conditions may be nondimensionalized in the same way, but they are not needed for the asymptotic reductions considered here.

\subsection{Ideal-gas specialization and conservative dimensionless form}

We consider a calorically perfect ideal gas with constant specific heats $c_p$ and $c_v$ at constant pressure and constant volume, respectively, satisfying
\begin{equation}
c_p>c_v>0,
\end{equation}
and corresponding specific gas constant
\begin{equation}
R\coloneqq c_p-c_v.
\end{equation}
To prepare for the subsequent nondimensionalization, we fix a thermodynamic reference state $(p_r,\vartheta_r)$ with $\vartheta_r>0$ and take the specific free enthalpy $\chi$ to be given by
\begin{equation}
\label{eq:free.enthalpy.ideal.gas}
\chi(p,\vartheta)
=R\vartheta\ln\frac{p}{p_r}
+c_p\vartheta\Big(1
-\ln\frac{\vartheta}{\vartheta_r}\Big).
\end{equation}
Specializing the thermodynamic relations \eqref{eq:CN.thermodynamic.relations} accordingly, we find that the mass density $\varrho$ and specific entropy $\eta$ are given by
\begin{equation}
\label{eq:ideal.gas.rho.eta}
\varrho=\frac{p}{R\vartheta}
\qquad\text{and}\qquad
\eta
=-R\ln\frac{p}{p_r}
+c_p\ln\frac{\vartheta}{\vartheta_r}.
\end{equation}
In particular, we require that the thermodynamic reference state be consistent with
\begin{equation}
\varrho_r=\frac{p_r}{R\vartheta_r}.
\end{equation}
Invoking \eqref{eq:free.enthalpy.def} and \eqref{eq:ideal.gas.rho.eta}, we recover the classical relation $\varepsilon=c_v\vartheta$ for the specific internal-energy, and, hence, by \eqref{eq:total.specific.energy}, the representation
\begin{equation}
e=c_v\vartheta+\tfrac12|\vl|^2
\end{equation}
for the specific total energy.

We next choose a characteristic velocity scale $U_r$ and a characteristic length scale $L$. The associated time scale is $L/U_r$. We also define reference values of the transport coefficients by evaluating the constitutive moduli at the reference thermodynamic state:
\begin{equation}
\mu_r\coloneqq\mu(p_r,\vartheta_r),
\qquad
\kappa_r\coloneqq\kappa(p_r,\vartheta_r),
\qquad
\iota_r\coloneqq\iota(p_r,\vartheta_r).
\end{equation}
To pass to a dimensionless formulation, we temporarily use a superscripted star to distinguish dimensional quantities. Accordingly, we set
\begin{equation}
\left\{
\begin{gathered}
x=\frac{x^\star}{L},
\qquad
t=\frac{U_r t^\star}{L},
\qquad
\varrho=\frac{\varrho^\star}{\varrho_r},
\qquad
\vel=\frac{\vel^\star}{U_r},
\qquad
\vl=\frac{\vl^\star}{U_r},
\qquad
\jl=\frac{\jl^\star}{U_r},
\\[4pt]
\vartheta=\frac{\vartheta^\star}{\vartheta_r},
\qquad
p=\frac{p^\star}{p_r},
\qquad
e=\frac{e^\star}{U_r^2},
\qquad
\bs{b}=\frac{\bs{b}^\star L}{\varrho_r U_r^2},
\qquad
r=\frac{r^\star L}{\varrho_r U_r^3},
\end{gathered}
\right.
\end{equation}
and
\begin{equation}
\Grad=L\mskip2mu\Grad^\star,
\qquad
\Div=L\mskip2mu\Div^\star,
\qquad
\mu=\frac{\mu^\star}{\mu_r},
\qquad
\zeta=\frac{\zeta^\star}{\mu_r},
\qquad
\kappa=\frac{\kappa^\star}{\kappa_r},
\qquad
\iota=\frac{\iota^\star}{\iota_r}.
\end{equation}
Let
\begin{equation}
a_r \coloneqq \sqrt{\gamma R \vartheta_r}
\end{equation}
denote the sound speed at the reference state. In addition to the conventional definitions
\begin{equation}
\mathrm{Ma}=\frac{U_r}{a_r},
\qquad
\mathrm{Re}=\frac{\varrho_r U_r L}{\mu_r},
\qquad\text{and}\qquad
\mathrm{Pr}=\frac{c_p\mu_r}{\kappa_r}
\end{equation}
of the Mach, Reynolds, and Prandtl numbers, we introduce the Brenner number
\begin{equation}
\mathrm{Br}=\frac{\iota_r\varrho_r U_r}{L},
\end{equation}
which is a measure of the strength of pressure-gradient-driven relative transport between mass and linear momentum, relative to the advective scale set by $U_r$ and $L$.

Upon suppressing the superscripted stars, we obtain the bulk equations in the conservative dimensionless form
\begin{equation}\label{eq:dimensionless.bulk}
\left\{
\begin{aligned}
& \partial_t\varrho+\Div(\varrho\vel)=0, \\[4pt]
& \partial_t(\varrho\vl)
+\Div\Big(\varrho\vl\otimes\vel+\frac{1}{\gamma\mathrm{Ma}^2}p\bs{1}\Big)
=\Div\Big(\frac{1}{\mathrm{Re}}\br{S}+\frac{1}{2}\varrho\vl\wedge\vel\Big)+\bs{b}, \\[4pt]
& \partial_t(\varrho e)
+\Div\Big(\Big(\varrho e+\frac{1}{\gamma\mathrm{Ma}^2}p\Big)\vel\Big) \\[4pt]
& \qquad
=\Div\Big(\Big(\frac{1}{\mathrm{Re}}\br{S}-\frac{1}{2}\varrho\vl\wedge\vel\Big)\vl
-\frac{1}{(\gamma-1)\mathrm{Re}\mathrm{Pr}\mathrm{Ma}^2}\bs{q}\Big)
+\frac{1}{2}(\varrho\vel\wedge\vl)\twovdots\Wl+\bs{b}\cdot\vl+r,
\end{aligned}
\right.
\end{equation}
To close the system \eqref{eq:dimensionless.bulk}, we recover the thermodynamic and kinematic quantities from the conservative variables through the relations
\begin{equation}\label{eq:dimensionless.constitutive}
\left\{
\begin{gathered}
\vartheta=\gamma(\gamma-1)\mathrm{Ma}^2\Big(e-\frac{1}{2}|\vl|^2\Big), \qquad
p=\varrho\vartheta, \\[4pt]
\vel=\vl-\frac{\mathrm{Br}}{\gamma\mathrm{Ma}^2}\iota(p,\vartheta)\Grad p, \qquad
\br{S}=2\mu(p,\vartheta)\Dl^0+\zeta(p,\vartheta)(\tr\Dl)\bs{1}, \qquad
\bs{q}=-\kappa(p,\vartheta)\Grad\vartheta.
\end{gathered}
\right.
\end{equation}

\subsection{Navier--Stokes--Fourier limit}

On passing to the limit $\mathrm{Br}\to0$ in \eqref{eq:dimensionless.constitutive}$_3$, the relative field $\jl=\vl-\vel$ vanishes. Accordingly, the specific linear momentum and the velocity coalesce, the antisymmetric term $\frac{1}{2}\varrho \vl \wedge \vel$ disappears from \eqref{eq:dimensionless.bulk}, and the system reduces to the classical compressible Navier--Stokes--Fourier equations. We therefore recover the classical model from the present theory if mass transport and linear-momentum transport are not distinguished.

\subsection{Distinguished low-Mach limit}

To obtain a nontrivial low-Mach limit, we let $\mathrm{Ma}\to 0$ while taking the Reynolds and Prandtl numbers $\mathrm{Re}$ and $\mathrm{Pr}$ both to be of order unity. We further assume that the Brenner number $\mathrm{Br}$ scales quadratically with $\mathrm{Ma}$:
\begin{equation}
\label{eq:scaling}
\mathrm{Br}=\mathcal{O}(\mathrm{Ma}^2).
\end{equation}
With \eqref{eq:scaling}, we define a distinguished low-Mach regime in which compressibility effects are balanced by pressure-gradient-driven relative transport. Accordingly, we assume that $\varrho$, $\vartheta$, $p$, $\vel$, and $\vl$ admit expansions of the form
\begin{equation}
\left\{
\begin{gathered}
\varrho = \varrho_0 + \mathrm{Ma}^2 \varrho_1 + o(\mathrm{Ma}^2),
\qquad
\vartheta = \vartheta_0 + \mathrm{Ma}^2 \vartheta_1 + o(\mathrm{Ma}^2),
\qquad
p = p_0 + \mathrm{Ma}^2 p_1 + o(\mathrm{Ma}^2),
\\[4pt]
\vel = \vel_0 + o(1),
\qquad\text{and}\qquad
\vl = {\vl}_0 + o(1),
\end{gathered}
\right.
\end{equation}
from \eqref{eq:dimensionless.constitutive}$_2$, we see that
\begin{equation}
p_0 = \varrho_0 \vartheta_0
\qquad\text{and}\qquad
p_1 = \varrho_0 \vartheta_1 + \varrho_1 \vartheta_0.
\end{equation}
Moreover, since, from the $\mathcal{O}(\mathrm{Ma}^{-2})$ terms of \eqref{eq:dimensionless.bulk}$_2$,  
\begin{equation}
\Grad p_0 = \bs{0},
\end{equation}
we see that $p_0$ must be spatially uniform. From the leading order contribution of \eqref{eq:dimensionless.bulk}$_3$, we thus obtain the solvability condition
\begin{equation}
\label{eq:low.mach.solvability}
\frac{1}{\gamma}\frac{\mathrm{d}p_0}{\mathrm{d}t}
+ p_0 \Div\vel_0
=
- \frac{1}{\mathrm{Re}\mathrm{Pr}} \Div\bs{q}_0,
\qquad
\bs{q}_0 = - \kappa(p_0,\vartheta_0)\Grad\vartheta_0.
\end{equation}

At leading order, the bulk equations reduce to
\begin{equation}
\label{eq:low.mach.leading.mass}
\partial_t \varrho_0 + \Div(\varrho_0 \vel_0) = 0
\end{equation}
and
\begin{equation}
\label{eq:low.mach.leading.momentum}
\partial_t(\varrho_0 {\vl}_0)
+ \Div(\varrho_0 {\vl}_0 \otimes \vel_0)
+ \frac{1}{\gamma}\Grad p_1
=
\Div\Big(\frac{1}{\mathrm{Re}} \br{S}_0 + \frac{1}{2}\varrho_0 {\vl}_0 \wedge \vel_0\Big)
+ \bs{b}_0,
\end{equation}
with
\begin{equation}
\br{S}_0
=
2\mu(p_0,\vartheta_0){\Dl}_0^{\mskip-8.5mu0}
+\zeta(p_0,\vartheta_0)(\tr{\Dl}_0)\bs{1}.
\end{equation}

Finally, \eqref{eq:dimensionless.constitutive}$_{2,3}$ lead to the condition
\begin{equation}
\label{eq:low.mach.relative.field}
{\vl}_0=\vel_0
+\frac{\mathrm{Br}}{\gamma\mathrm{Ma}^2}
\iota(p_0,\vartheta_0)\Grad p_1.
\end{equation}
Thus, under the scaling \eqref{eq:scaling}, the low-Mach limit retains distinct mass and linear-momentum transport at leading order. If, in addition, $\mathrm{Br}/\mathrm{Ma}^2\to0$, then \eqref{eq:low.mach.relative.field} leads to the condition $\jl_0={\vl}_0-\vel_0=\bs{0}$ and we recover the classical low Mach number Navier--Stokes--Fourier regime.

\section{Summary}
\label{sec:summary}

We developed a thermodynamically consistent continuum theory for compressible, viscous, heat-conduct\-ing fluids in which the velocity entering the balance of mass is distinguished from the specific linear momentum entering the balances of linear momentum and total energy. In the mechanical formulation, the traction-power term and the kinetic-energy density are both expressed in terms of the specific linear momentum. By enforcing angular-momentum balance, we found that once the velocity and the specific linear momentum are allowed to differ, the Cauchy stress need not be symmetric. However, the classical symmetric stress is recovered if the mismatch between the fields governing mass transport and momentum transport vanishes.

We then enforced the second law and derived the accompanying thermodynamic restrictions. In particular, we identified admissible forms of the internal body-couple and internal-energy flux, reduced the residual dissipation to mechanical, thermal, and pressure-gradient contributions, and obtained a simple class of admissible constitutive laws with nonnegative transport moduli. In contrast with Brenner's \cite{Bre06} mass- and volume-velocity theory, in which relative transport is taken proportional to the gradient of the logarithm of the mass density, we obtained a closure in which the relative transport is taken to be proportional to the pressure gradient. In the Supplementary Material, we show that this distinction is essential: once either angular-momentum balance or the Clausius--Duhem inequality is imposed, Brenner's theory collapses to the classical Navier--Stokes--Fourier theory.

We also extended the formulation to shocks and solid boundaries. Across shocks, we derived a free-enthalpy imbalance that separates mechanical and thermal contributions to interfacial dissipation. At a rigid, impermeable wall undergoing prescribed rigid motion, we reduced the wall dissipation inequality to a normal mechanical contribution together with thermal exchange, and on that basis we formulated simple admissible wall laws for both temperature-controlled and heat-flow-controlled settings.

For ideal gases we specialized the theory to a conservative dimensionless form and introduced the Brenner number as a measure of the strength of pressure-gradient-driven relative transport. Using this form, we identified two informative asymptotic regimes. When the Brenner number tends to zero, the relative transport disappears and the classical compressible Navier--Stokes--Fourier system is recovered. In a distinguished low-Mach regime, by contrast, the velocity governing mass transport and the specific linear momentum remain distinct at leading order. By analyzing these asymptotic regimes, we showed exactly how the present theory relates to Navier--Stokes--Fourier: the classical equations are recovered when relative transport is absent, whereas in a distinguished low-Mach regime relevant to motions of compressible fluids with strong density gradients, distinct mass and linear-momentum transport persists at leading order.

%-------------------------------------------------------------------------------%

\section*{Acknowledgements}

Luis Espath acknowledges support from the EPSRC Impact Acceleration Account (IAA) at the University of Nottingham. The authors also thank Space Forge for valuable discussions related to high-speed and non-equilibrium flow applications. Eliot Fried expresses gratitude for support provided by the Okinawa Institute of Science and Technology, funded by the Cabinet Office of the Government of Japan.

\section*{Declarations}

\medskip\noindent\textbf{Author contributions} All authors contributed equally to the conception, analysis, and writing of this work.

\medskip\noindent\textbf{Funding} This work was supported by the EPSRC Impact Acceleration Account (IAA).

\medskip\noindent\textbf{Data availability} No datasets were generated or analysed during the current study.

\medskip\noindent\textbf{Ethical approval} Not applicable.

\medskip\noindent\textbf{Competing interests} The authors declare no competing interests.

%-------------------------------------------------------------------------------%

\appendix

\section{Transport and divergence identities}
\label{sec:identities}

Consider a fixed control volume $\prt$ partitioned by a shock surface $\srf_t$ into complementary open subregions $\prt^+$ and $\prt^-$. Let $\bs{m}$ denote the unit normal to $\srf_t$ directed from $\prt^-$ into $\prt^+$, and let $\xi$ denote the scalar normal velocity of $\srf_t$. 

Suppose that $f$ and $\bs{g}$ are scalar and vector fields, respectively, that are smooth on $\prt^\pm$ but suffer finite jumps across $\srf_t$. Denoting by $f^\pm$ and $\bs{g}^\pm$ the restrictions of $f$ and $\bs{g}$ to $\srf_t$, and writing
\begin{equation}
\jump{f}=f^+-f^-
\qquad\text{and}\qquad
\jump{\bs{g}}=\bs{g}^+-\bs{g}^-,
\end{equation}
we find that
\begin{equation}
\label{eq:ui1}
\ddt\intP f\dv=\intP\frac{\partial f}{\partial t}\dv
+\int\limits_{\prt\cap\srf_t}\jump{f}\xi\da_t,
\qquad
\intdP\bs{g}\cdot\bs{n}\da
=\intP\Div\bs{g}\dv
+\int\limits_{\prt\cap\srf_t}\jump{\bs{g}}\cdot\bs{m}\da_t,
\end{equation}
where $\text{d}a_t$ denotes the elemental area on $\srf_t$. Setting $f=\varrho$ in \eqref{eq:ui1}$_1$ and $\bs{g}=\varrho\vel$ in \eqref{eq:ui1}$_2$ and using the definition of the material time derivative, we may rewrite the mass balance \eqref{eq:mass.balance.gen} for $\prt$ in the form
\begin{equation}
\label{eq:mass.balance.P.general}
\intP\Big(\Dt{\varrho}+\varrho\mskip2mu\Div\vel\Big)\dv
+\int\limits_{\prt\cap\srf_t}\jump{\varrho(\xi-\vel\cdot\bs{m})}\da_t=0.
\end{equation}
Choosing $\prt$ to be compactly supported about a point away from $\srf_t$, so that $\prt\cap\srf_t=\varnothing$, we may localize \eqref{eq:mass.balance.P.general} to obtain the field equation expressing the pointwise balance of mass in $\prt^-\cup\prt^+$:
\begin{equation}
\label{eq:mass.balance.field.equation.B}
\Dt{\varrho}+\varrho\mskip2mu\Div\vel=0.
\end{equation}
Substituting \eqref{eq:mass.balance.field.equation.B} back into \eqref{eq:mass.balance.P.general}, and recalling that $\prt=\prt^-\cup\prt^+$ with $\prt^\pm$ separated by $\prt\cap\srf_t$, we find that the volume integral vanishes identically and that \eqref{eq:mass.balance.P.general} reduces to the interfacial statement
\begin{equation}
\label{eq:mass.balance.P.reduced}
\int\limits_{\prt\cap\srf_t}\jump{\varrho(\xi-\vel\cdot\bs{m})}\da_t=0.
\end{equation}
Localizing \eqref{eq:mass.balance.P.reduced} about an interior point of $\srf_t$, so that $\prt\cap\srf_t$ is an arbitrary surface patch containing that point, yields the jump condition expressing mass balance on $\srf_t$:
\begin{equation}
\label{eq:mass.balance.jump.condition.B}
\jump{\varrho(\xi-\vel\cdot\bs{m})}=0.
\end{equation}
In view of \eqref{eq:mass.balance.jump.condition.B}, the quantity $\varrho(\xi-\vel\cdot\bs{m})$ has equal
one-sided limits on $\srf_t$. This common value represents the normal mass flux relative to
the shock. We therefore introduce a scalar field $\jmath$ on $\srf_t$ to denote that flux:
\begin{equation}
\label{eq:mass.flux}
\jmath:=\varrho^+(\xi-\vel^+\cdot\bs{m})
=\varrho^-(\xi-\vel^-\cdot\bs{m}).
\end{equation}

From the definition of the material time derivative and \eqref{eq:mass.balance.field.equation.B}, it follows that, for any sufficiently smooth field $\varphi$,
\begin{equation}
\label{eq:ui3}
\frac{\partial(\varrho\varphi)}{\partial t}=\Dt{(\varrho\varphi)}-\vel\cdot\Grad(\varrho\varphi)
=\varrho\Dt\varphi-\Div(\varrho\varphi\vel)+\Big(\Dt\varrho+\varrho\mskip2mu\Div\vel\Big)\varphi
=\varrho\Dt\varphi-\Div(\varrho\varphi\vel)
\end{equation}
on $\prt^-\cup\prt^+$. Allowing for the possibility that $\varphi$ may suffer a finite jump discontinuity across $\srf_t$, we integrate \eqref{eq:ui3} over $\prt$ and use \eqref{eq:ui1}$_1$ with $f=\varrho\varphi$ and \eqref{eq:ui1}$_2$ with $\bs{g}=\varrho\varphi\vel$ to obtain the transport relation
\begin{equation}
\label{eq:reynolds.transport.shock}
\ddt\intP\varrho\varphi\dv=\intP\varrho\Dt\varphi\dv
-\intdP\varrho\varphi\vel\cdot\bs{n}\da
-\int\limits_{\prt\cap\srf_t}\jmath\jump{\varphi}\da_t,
\end{equation}
which holds irrespective of the tensorial order of $\varphi$. 

For a second-order tensor field $\br{A}$ that is smooth on $\prt^-\cup\prt^+$ and admits a finite jump across $\srf_t$, the counterpart of \eqref{eq:ui1}$_2$ reads
\begin{equation}
\label{eq:ui2}
\intdP\br{A}\bs{n}\da
=\intP\Div\br{A}\dv
+\int\limits_{\prt\cap\srf_t}\jump{\br{A}}\bs{m}\da_t.
\end{equation}
Since, for any fixed vector $\bs{c}$,
\begin{equation}
\Div((\bs{g}\cdot\bs{c})\br{A})
=(\Div\br{A}\otimes\bs{g}+\br{A}(\Grad\bs{g})^{\trans\mskip3mu})\bs{c},
\end{equation}
we see from \eqref{eq:ui2} that if $\bs{g}$ is smooth on $\prt$ and $\br{A}$ is smooth on $\prt^-\cup\prt^+$ but suffers a finite jump discontinuity across $\srf_t$, then
\begin{equation}
\Big(\intdP\br{A}\bs{n}\otimes\bs{g}\da\Big)\bs{c}
=\intdP(\bs{g}\cdot\bs{c})\br{A}\bs{n}\da
=\Big(\intP(\Div\br{A}\otimes\bs{g}+\br{A}(\Grad\bs{g})^{\trans\mskip3mu})\dv
+\int\limits_{\prt\cap\srf_t}\bs{g}\wedge\jump{\br{A}}\bs{m}\da_t\Big)\bs{c}
\end{equation}
and, thus, with reference to the definition \eqref{eq:wedge.product} of the wedge product, that
\begin{equation}
\label{eq:ui7}
\intdP\bs{g}\wedge\br{A}\bs{n}\da
=\intP(\bs{g}\wedge\Div\br{A}
+(\Grad\bs{g})\br{A}^{\mskip-2mu\trans}-\br{A}(\Grad\bs{g})^{\trans\mskip2mu})\dv
+\int\limits_{\prt\cap\srf_t}
\bs{g}\wedge\jump{\br{A}}\bs{m}\da_t.
\end{equation}

%-------------------------------------------------------------------------------%

% \clearpage
\makeatletter
\let\sectionname\@empty
\makeatother
\footnotesize

% \bibliographystyle{unsrt}
% \bibliography{bib}

\clearpage
\makeatletter
\let\sectionname\@empty
\gdef\theHsection{\arabic{section}}
\gdef\theHsubsection{supplement.\arabic{section}.\arabic{subsection}}
\gdef\theHequation{supplement.\arabic{equation}}
\xdef\Hy@chapapp{supplement}
\makeatother
\section*{Supplementary Material}
\setcounter{section}{0}
\setcounter{subsection}{0}
\setcounter{equation}{0}
\renewcommand{\thesection}{S\arabic{section}}
\renewcommand{\theequation}{S\arabic{equation}}

\section{Introduction}

Here, we show that Brenner's~\cite{Bre06SI} theory reduces to the Navier--Stokes--Fourier theory if either balance of angular momentum or the second law of thermodynamics is imposed in addition to his governing equations. To place this assertion in context, the admissibility of dissipative mass flux in continuum hydrodynamics has been examined from two rather different perspectives. \"Ottinger, Struchtrup and Liu~\cite{OSL} analyzed such fluxes in a general continuum framework and identified strong restrictions arising from angular momentum and other mechanical principles, ultimately ruling them out in the hydrodynamic regime. By contrast, Dadzie and Reese~\cite{DR} argued that certain apparent inconsistencies of volume-mass-diffusion models can be resolved in a scale-separated, high-local-Knudsen setting and exhibited a distinct mechanically consistent model. The purpose of the present note is narrower and more specific: it addresses Brenner's later mass-/volume-velocity theory as written and shows that, with his governing equations and constitutive prescriptions, the theory reduces to Navier--Stokes--Fourier if either balance of angular momentum or the Clausius--Duhem inequality is imposed.

\section{Remarks on Brenner's mass- and volume-velocity theory}
\label{sec:brenner}

If external body forces and supplies of heat are neglected, the local balance equations for mass, linear momentum, and energy of Brenner's \cite{Bre06SI} theory can be expressed as
\begin{equation}\label{eq:brenner.mass}
\partial_t\varrho+\Div(\varrho\vm)=0,
\end{equation}
\begin{equation}\label{eq:brenner.momentum}
\partial_t(\varrho\vv)+\Div(\varrho\vv\otimes\vm)-\Div(\br{S}-p\bs{1})=\bs{0},
\end{equation}
and
\begin{equation}\label{eq:brenner.energy}
\partial_t(\varrho e)+\Div(\varrho e\vm)+\Div(\bs{q}-p\jv)-\Div((\br{S}-p\bs{1})^{\trans}\vv)=0,
\end{equation}
where the fields $\vm$ and $\vv$ respectively represent the spatial description of the mass- and volume-average velocities, $\partial_t$ is the partial time-derivative, div is the spatial divergence operator, $\br{S}$ is the viscous part of the Cauchy stress, $p$ is the thermodynamic pressure, $e$ is the specific total energy defined by $e\coloneqq\varepsilon+\fr{1}{2}|\vv|^{2}$ with $\varepsilon$ the specific internal energy, $\bs{q}$ is the heat flux, and
\begin{equation}\label{eq:dfn.jv}
\jv\coloneqq\vv-\vm,
\end{equation}
is the diffusive volume flux.

To obtain a closed system of evolution equations, the balances \eqref{eq:brenner.mass}--\eqref{eq:brenner.energy} are augmented by constitutive relations for $\varepsilon$, $p$, $\br{S}$, $\bs{q}$, and $\jv$. Beginning with thermodynamic specifications, let $\eta$ denote the specific entropy and assume that the specific free enthalpy $\chi\coloneqq\varepsilon-\vartheta\eta+p/\varrho$ is determined by a response function $\hat\chi$ depending on $p$ and $\vartheta$:
\begin{equation}
\label{eq:thermo.chi}
\chi=\hat\chi(p,\vartheta).
\end{equation}
Then, $\varrho$ and $\eta$ determined through $\hat\chi$ by
\begin{equation}\label{eq:thermo.eps}
\hat\varrho(p,\vartheta)=\frac{1}{\partial_p\hat\chi(p,\vartheta)}
\qquad\text{and}\qquad
\hat\eta(p,\vartheta)=-\partial_\vartheta\hat\chi(p,\vartheta).
\end{equation}
Next, with the definitions
\begin{equation}\label{eq:kinematics.Dv}
\Dv\coloneqq\fr{1}{2}(\Grad\vv+(\Grad\vv)^{\trans})
\qquad\text{and}\qquad
\Dv^{0}\coloneqq\Dv-\tfrac13(\tr\Dv)\bs{1}
\end{equation}
of the stretching tensor and its deviatoric part, the viscous part of the Cauchy stress is taken to have the linear, isotropic form
\begin{equation}\label{eq:const.S}
\br{S}=2\mu(p,\vartheta)\Dv^{0}+\zeta(p,\vartheta)(\tr\Dv)\bs{1},
\end{equation}
where $\mu\ge0$ and $\zeta\ge0$ are the shear and bulk viscosities. The heat flux is assumed to obey Fourier's law
\begin{equation}\label{eq:const.q}
\bs{q}=-\kappa(p,\vartheta) \mskip2mu \Grad\vartheta,
\end{equation}
where $\kappa\ge0$ is the thermal conductivity. Finally, the diffusive volume flux is taken to be
\begin{equation}\label{eq:const.jv}
\jv=\iota(p,\vartheta) \mskip2mu \Grad\ln\varrho,
\end{equation}
where $\iota\ge0$ is a phenomenological modulus. Combining \eqref{eq:const.q} and \eqref{eq:const.jv} leads to an ancillary constitutive relation for the total energy flux,
\begin{equation}\label{eq:ju.explicit}
\ju\coloneqq\bs{q}-p\jv=-\kappa(p,\vartheta) \mskip2mu \Grad\vartheta-p\iota(p,\vartheta) \mskip2mu \Grad\varrho.
\end{equation}

On a fixed solid wall, modeled as rigid and impermeable, the boundary conditions consist of
\begin{equation}
\label{eq:brenner.dprt}
(\bs{1}-\bs{n}\otimes\bs{n})\vv=\bs{0}
\qquad\text{and}\qquad
\vm\cdot\bs{n}=0.
\end{equation}
Since, by \eqref{eq:dfn.jv} and \eqref{eq:brenner.dprt}, $(\jv-\vv)\cdot\bs{n}=0$ and $(\bs{1}-\bs{n}\otimes\bs{n})(\jv+\vm)=\bs{0}$, it follows from \eqref{eq:const.jv} that the remaining components of $\vv$ and $\vm$ on the wall must satisfy
\begin{equation}
\label{eq:brenner.dprtderived}
(\bs{1}-\bs{n}\otimes\bs{n})\vm=-\iota(p,\vartheta) \Grads\ln\varrho
\qquad\text{and}\qquad
\vv\cdot\bs{n}=\iota(p,\vartheta)\partial_n\ln \varrho,
\end{equation}
with $\Grads$ and $\partial_n$ being the surface gradient and the normal derivative on the wall, respectively.

\subsection{Balance of angular momentum}

Brenner \cite{Bre06SI} does not formulate a local balance of angular momentum or a local statement of the second law.
Thus, to examine balance of angular momentum, consider the identities
\begin{equation}
\label{eq:wedge_identities}
\Grad\bs{r}=\bs{1},
\qquad
\bs{r}\wedge\partial_t\bs{w}=\partial_t(\bs{r}\wedge\bs{w}),
\qquad\text{and}\qquad
\bs{r}\wedge\Div\br{T}=\Div(\bs{r}\wedge\br{T})+\br{T}-\br{T}^{\trans},
\end{equation}
we see from the balance \eqref{eq:brenner.momentum} of linear momentum and the consequence $\br{S}=\br{S}^{\trans}$ of \eqref{eq:const.S} that
\begin{align}
\bs{0}
&=\bs{r}\wedge(\partial_t(\varrho\vv)+\Div(\varrho\vv\otimes\vm)-\Div(\br{S}-p\bs{1}))
\notag\\[4pt]
&=\partial_t(\varrho\bs{r}\wedge\vv)
+\Div(\bs{r}\wedge(\varrho\vv\otimes\vm-(\br{S}-p\bs{1})))
+\varrho\vv\wedge\vm,
\label{eq:AM.tensor.refined}
\end{align}
where $\bs{r}\wedge\br{T}$ is the third-order tensor defined such that, for any vector $\bs{c}$,
\begin{equation}
(\bs{r}\wedge\br{T})\br{c}=\bs{r}\otimes\br{T}\bs{c}-\br{T}\bs{c}\otimes\bs{r}.
\end{equation}
The first two terms on the second line of \eqref{eq:AM.tensor.refined} represent, respectively, the local time rate and the flux of the moment of the linear momentum, and the additional term $\varrho\vv\wedge\vm$ corresponds to a production of angular momentum. Thus, in the absence of body couples and couple stresses, the balance of angular momentum holds if and only if $\vv$ and $\vm$ satisfy
\begin{equation}\label{eq:collinearity}
\vv\wedge\vm=\bs{0},
\end{equation}
and, consequently, are locally collinear. Thus, $\vv=\lambda\vm$ for some scalar field $\lambda$. Using this condition in \eqref{eq:dfn.jv}, we see that $\jv=(\lambda-1)\vm$ and, by \eqref{eq:const.jv}, that
\begin{equation}
(\lambda-1)\vm=\iota(p,\vartheta) \mskip2mu \Grad\ln\varrho.
\end{equation}
Thus, unless $\lambda\equiv1$ or $\iota\equiv0$, $\Grad\varrho$ must be locally parallel to $\vm$. Imposing such an alignment for general processes and boundary data is, however, untenable. The natural alternative is to require that $\iota\equiv0$, in which case $\jv\equiv\bs{0}$ and, hence, $\vv\equiv\vm$. Under this reduction, the balance of angular momentum reduces to the symmetry of $\br{S}$, which is ensured by \eqref{eq:const.S}. In the formal case $\lambda=-1$, $\vv\wedge\vm=\bs{0}$ also holds, but then $\jv=-2\vm$ and the same alignment constraint arises. On setting $\iota\equiv0$, this constraint vanishes and it again follows that $\vv\equiv\vm$.

Related angular-momentum objections to dissipative mass flux were identified in a broader hydrodynamic setting by \"Ottinger, Struchtrup, and Liu \cite{OSL}. For Brenner's \cite{Bre06SI} theory, the obstruction appears directly in the local moment balance and, in conjunction with the constitutive relation \eqref{eq:const.jv} for the diffusive volume flux, forces collapse to the classical theory.

\subsection{Second law of thermodynamics}

Turning to the second law, we form the scalar product of the balance \eqref{eq:brenner.momentum} of linear momentum with $\vv$ and use the balance of mass \eqref{eq:brenner.mass} to consolidate the advective terms, yielding the kinetic-energy identity
\begin{equation}
\label{eq:ke.identity}
\partial_t(\fr{1}{2}\varrho|\vv|^2)
+\Div(\fr{1}{2}\varrho|\vv|^2\,\vm)
-\Div((\br{S}-p\bs{1})^{\trans}\vv)
+(\br{S}-p\bs{1})\twovdots\Grad\vv=0.
\end{equation}
Next, we subtract \eqref{eq:ke.identity} from the balance \eqref{eq:brenner.energy} of energy to obtain the balance of internal energy
\begin{equation}
\label{eq:int.energy.balance}
\varrho\dot{\varepsilon}+\Div(\bs{q}-p\jv)
=(\br{S}-p\bs{1})\twovdots\Grad\vv.
\end{equation}
We impose the second law via the Clausius--Duhem inequality, which, recalling that external supplies of heat have been neglected, reads
\begin{equation}
\label{eq:CD.local.again}
\partial_t(\varrho\eta)+\Div(\varrho\eta\vm)
+\Div\Big(\frac{\bs{q}}{\vartheta}\Big)\ge0.
\end{equation}
Subtracting the product of \eqref{eq:CD.local.again} and $\vartheta$ from \eqref{eq:int.energy.balance} and using the definition $\chi\coloneqq\varepsilon-\vartheta\eta+p/\varrho$ of the specific free enthalpy, we arrive at the inequality
\begin{equation}\label{eq:free.energy.form.again}
\varrho\Big(\Dt{\chi}-\frac{1}{\varrho}\Dt{p}
+\eta\Dt{\vartheta}\Big)
-\br{S}\twovdots\Dv
-\jv\cdot\Grad p
+\bs{q}\cdot\Grad\ln\vartheta\ge0.
\end{equation}
Inserting \eqref{eq:thermo.chi}, \eqref{eq:thermo.eps}, and \eqref{eq:const.S}--\eqref{eq:const.jv} in \eqref{eq:free.energy.form.again}, we obtain the reduced dissipation inequality
\begin{equation}\label{eq:diss.specialized}
2\mu(p,\vartheta)|\Dv^{0}|^2
+\zeta(p,\vartheta)(\tr\Dv)^2
+\kappa(p,\vartheta)\vartheta|\Grad\ln\vartheta|^2
+\iota(p,\vartheta) \mskip2mu \Grad p\cdot\Grad\ln\varrho\ge0.
\end{equation}
The first three terms on the left-hand side of \eqref{eq:diss.specialized} coincide with the expression for the dissipation in the classical Navier--Stokes--Fourier theory. The remaining pressure-diffusion term $\iota(p,\vartheta) \mskip2mu \Grad p\cdot\Grad\ln\varrho$ has no fixed sign for general processes and boundary data. Absent supplementary restrictions or compensating cross-effects, it follows from \eqref{eq:diss.specialized} that $\iota$ must satisfy $\iota\equiv0$. Under this condition $\jv\equiv\bs{0}$, $\vv\equiv\vm$, and the dissipation reduces to its classical form.

\"Ottinger, Struchtrup and Liu \cite{OSL} discuss the admissibility of dissipative mass flux in a broader hydrodynamic setting. For Brenner's \cite{Bre06SI} theory, the obstruction appears directly in the
local Clausius--Duhem inequality and, in conjunction with the constitutive relation \eqref{eq:const.jv} for the diffusive volume flux, forces collapse to the classical theory.

\subsection{Wall conditions}

On a fixed solid wall, modeled as rigid and impermeable, the implications $\iota\equiv0$ and $\vv\equiv\vm$ force the right-hand sides of \eqref{eq:brenner.dprtderived} to vanish. With the identification $\bs{\upsilon}=\vv=\vm$, \eqref{eq:brenner.dprt} and \eqref{eq:brenner.dprtderived} lead to the classical wall conditions
\begin{equation}
\label{eq:wall.reduced}
(\bs{1}-\bs{n}\otimes\bs{n})\bs{\upsilon}=\bs{0}
\qquad\text{and}\qquad
\bs{\upsilon}\cdot\bs{n}=0.
\end{equation}

\section{Summary}

The foregoing analysis yields a model-specific collapse result for Brenner's \cite{Bre06SI} mass- and volume-velocity theory. If either local balance of angular momentum or the Clausius--Duhem inequality is imposed, in addition to Brenner's governing equations and constitutive prescriptions, then the phenomenological modulus $\iota$ in \eqref{eq:const.jv} must vanish. Consequently, $\jv\equiv\bs{0}$, the two kinematic fields coincide, $\vv\equiv\vm$, the balance laws reduce to the classical Navier--Stokes--Fourier system, and the wall conditions reduce to the classical no-slip/no-penetration conditions \eqref{eq:wall.reduced}.

The two routes to this conclusion are independent. The local moment balance associated with \eqref{eq:brenner.momentum} yields the collinearity condition \eqref{eq:collinearity}; together with \eqref{eq:const.jv}, this would otherwise require a non-generic alignment between $\Grad\varrho$ and $\vm$, which is untenable for general processes and boundary data. The reduced dissipation inequality \eqref{eq:diss.specialized}, on the other hand, contains the term $\iota(p,\vartheta)\mskip2mu\Grad p\cdot\Grad\ln\varrho$, which has no definite sign; absent additional constitutive couplings or supplementary restrictions, thermodynamic admissibility therefore again requires $\iota\equiv0$. Thus, for Brenner's \cite{Bre06SI} theory, either angular-momentum balance or the second law alone is sufficient to force collapse to the classical theory. This conclusion complements the broader admissibility analysis of \"Ottinger, Struchtrup and Liu~\cite{OSL} and does not address the structurally different scale-separated, high-local-Knudsen models considered by Dadzie and Reese~\cite{DR}.

% \clearpage
\renewcommand{\refname}{Supplementary References}

\end{document}